\documentclass[showpacs,prd,draft,
preprint,tightenlines,
nofootinbib,11pt]{revtex4}

\usepackage{bm}
\usepackage{epsf}
\usepackage{dcolumn}

\newcommand{\beq}{\begin{equation}}
\newcommand{\eeq}{\end{equation}}
\newcommand{\bea}{\begin{eqnarray}}
\newcommand{\eea}{\end{eqnarray}}

\newcommand{\mc}[3]{\multicolumn{#1}{#2}{#3}}
\newcommand{\order}[1]{O\left(#1\right)}  
\newcommand{\wtilde}[1]{\widetilde{#1}}
\newcommand{\prob}{\mathcal{P}}

\newcommand{\VEC}[1]{{\bf \bm{#1}}} 
\newcommand{\pslash}[1]{\rlap{/}\kern-0.8pt #1}
\newcommand{\Dslash}{\rlap{/}\kern-2.0pt D}

\newcommand{\tr}{{\rm Tr}}
\def\today{\number\day\space\ifcase\month\or
January\or February\or March\or April\or May\or June\or
July\or August\or September\or October\or November\or December\fi
\space\number\year}
\newcount\mins \newcount\hours
\def\now{\hours=\time \mins=\time
	\divide\hours by60 \multiply\hours by60 \advance\mins by-\hours
	\divide\hours by60 
	\number\hours:\ifnum\mins<10 0\fi\number\mins}

\def\stacksymbols #1#2#3#4{\def\theguybelow{#2}
    \def\verticalposition{\lower#3pt}
    \def\spacingwithinsymbol{\baselineskip0pt\lineskip#4pt}
    \mathrel{\mathpalette\intermediary#1}}
\def\intermediary #1#2{\verticalposition\vbox{\spacingwithinsymbol
    \everycr={}\tabskip0pt
    \halign{$\mathsurround0pt#1\hfil##\hfil$\crcr#2\crcr
             \theguybelow\crcr}}}

\newcommand{\psid}{\overline{\psi}}
\newcommand{\qd}{\overline{q}}
\newcommand{\psiud}{\overline{\Psi}}
\newcommand{\phiud}{\overline{\Phi}}
\newcommand{\hpsid}{\overline{\Psi}_H}

\newcommand{\nl}{\nonumber \\}
\newcommand{\wcal}{{\cal W}}
\newcommand{\xv}{\VEC{x}}

\newcommand{\pv}{\VEC{p}}
\newcommand{\piv}{\VEC{\pi}}
\newcommand{\kv}{\VEC{k}}
\newcommand{\ktint}{\int^{\pi/a}_{-\pi/a}\frac{dk_0}{2 \pi}}
\newcommand{\hktint}{\int^{\pi/2a}_{-\pi/2a}\frac{dk_0}{2 \pi}}

\newcommand{\ainv}{$a^{-1}$}

\newcommand{\tnaik}{y_{t,{\rm naik}}}
\newcommand{\chisqaug}{\chi^2_{\rm aug}}
\newcommand{\chisqaugdof}{\chisqaug/{\rm DoF}}

\newcommand{\act}{{\cal S}}

\newcommand{\delv}{{\bf \nabla}}
\newcommand{\delvt}{{\bf \tilde{\nabla}}}

\newcommand{\delfour}{{\Delta^{(4)}}}
\newcommand{\delsq}{\Delta^{(2)}}

\newcommand{\vev}[1]{\langle #1 \rangle}
\newcommand{\Mbz}{{(a_sM_0)}}
\newcommand{\sigmav}{\mbox{\boldmath$\sigma$}}

\newcommand{\Ev}{{\bf \tilde{E}}}
\newcommand{\Bv}{{\bf \tilde{B}}}
\newcommand{\half}{{\mbox{$\frac{1}{2}$}}}

\begin{document}


\title{Heavy-light mesons with staggered light quarks}

\author{Matthew Wingate} 
\author{Junko Shigemitsu}
\affiliation{Department of Physics,
The Ohio State University, Columbus, OH 43210, USA }
\author{Christine T.\ H.\ Davies}
\affiliation{Department of Physics \& Astronomy,
University of Glasgow, Glasgow, G12 8QQ, UK}
\author{G.\ Peter Lepage} 
\affiliation{ Newman Laboratory of Nuclear Studies,
Cornell University, Ithaca, NY 14853, USA}
\author{Howard D.\ Trottier}
\affiliation{Physics Department, Simon Fraser University,
Burnaby, B.C., V5A 1S6, Canada}

\date{\now\space on\space\today}

\begin{abstract}
 We demonstrate the viability of improved staggered light quarks 
in studies of heavy-light systems.  Our method 
for constructing heavy-light operators  
exploits the close relation between naive and staggered fermions.
The new approach is tested on quenched 
configurations using several staggered actions
combined with nonrelativistic heavy quarks.
Exploratory calculations of the $B_s$ meson kinetic mass, 
the hyperfine and $1P-1S$ splittings in $B_s$, and the decay constant 
$f_{B_s}$ are presented and compared to previous quenched 
lattice studies.  An important technical detail,
Bayesian curve-fitting, is discussed at length.
\end{abstract}

\preprint{OHSTPY-HEP-T-02-012}

\maketitle

\newpage


\section{Introduction}
\label{sec:intro}

Precise calculations of hadronic matrix elements are important
ingredients in the quest to constrain the flavor-mixing parameters of
the Standard Model, the Cabibbo-Kobayashi-Maskawa matrix
elements $V_{ff'}$.
For example, the main theoretical input in extracting the
ratio $|V_{td}/V_{ts}|$ involves a combination
of the decay constants, $f_B$ and $f_{B_s}$, which parameterize leptonic
$B$ and $B_s$ decays, and of the neutral $B$ and $B_s$ mixing parameters,
$B_B$ and $B_{B_s}$.  Uncertainties in these quantities, or more 
 specifically in the combination $\xi \equiv
(f_{B_s}/f_B)\sqrt{B_{B_s}/B_B}$, currently restrict our ability to 
carry out stringent consistency checks of the Standard Model
(e.g.\ see \cite{Ligeti:2001an}).
 If these theoretical errors could be reduced by a factor of 2 or more 
the impact would be immediate and far-reaching.  
Similarly, high precision theoretical calculations of form factors
governing $B\to D\ell\bar{\nu}$ and $B\to \pi\ell\bar{\nu}$ decays
are crucial to determinations of $|V_{cb}|$ and $|V_{ub}|$, respectively.

Monte Carlo simulation of QCD on a lattice will ultimately provide
the most accurate theoretical determinations of mixing parameters, 
decay constants, and form factors  since lattice QCD is 
one of the few systematically improvable approaches to QCD.  
 Understanding and removing systematic uncertainties  
in lattice calculations, however, is arduous and complicated, and much of the 
effort in lattice gauge theory over the past decade has focused on this task. 
One very promising outcome of all this activity is the emergence of
 improved staggered actions for light quarks combined with 
highly improved glue actions.  The MILC collaboration, for instance, 
works with the ``AsqTad'' quark action which is free from the 
leading discretization errors, 
including those arising from the breaking of the 
fermion doubling symmetry, so that the action is accurate up to
$\order{\alpha_s a^2}$ errors.  They employ the 
one-loop Symanzik improved glue action with errors coming in 
only at $\order{\alpha_s^2 a^2}$.  Staggered actions have an 
exact chiral symmetry at zero mass 
 and are much cheaper to simulate than Wilson-type 
quark actions, so it has been possible for the MILC collaboration 
to carry out unquenched simulations with much smaller dynamical 
quark masses than has been attempted in the past.  They are now 
starting to obtain impressive results for light hadron spectroscopy 
and light meson decay constants \cite{Bernard:2001av,Bernard:2002bk}.

In this article we demonstrate that improved staggered quarks can also be 
used very  effectively to simulate the light quark in heavy-light systems 
such as in $B$-physics.  The past decade has seen significant progress
in our ability to simulate heavy quarks accurately 
(the commonly used NRQCD action,
for instance, has errors coming in at 
$\order{\alpha_s a^2}$ and $\order{\alpha_s \Lambda_{\rm QCD}/M}$, and 
work is underway to remove the latter).  On the other hand, only 
Wilson-type actions have been used for the valence light quarks in
heavy-light mesons, baryons, and electroweak currents, making it difficult 
to go much below $m_{\rm strange}/2$ in the light quark mass due 
to the necessary computational expense.  Consequently, the extrapolation 
of simulation results to the chiral limit
leads to the dominant systematic error in studies of $B$ and 
$D$ mesons (aside from quenching uncertainties).  
Furthermore the leading discretization errors in heavy-light 
simulations come from the light quark sector since Wilson-type 
actions have worse finite lattice spacing errors than 
improved glue, improved staggered, or NRQCD actions.
This situation motivated us to initiate a new approach 
to heavy-light simulations, namely the use of improved staggered 
light quarks combined with nonrelativistic heavy quarks.  Our approach can 
trivially be modified to use a Wilson-like action for the
heavy quark instead of NRQCD.
The goal is to simulate $B$ physics at much smaller light quark masses 
than has been possible in the past and significantly reduce 
chiral extrapolation errors in decay constants, form factors and 
mixing parameters.  Work toward this goal has already started on the MILC 
dynamical configurations \cite{Wingate:2002ry}.  
It is important, however, to first establish 
that we understand how to combine staggered light and NRQCD or 
Wilson heavy fermions to form heavy-light operators, that we are 
able to carry out sophisticated fits to simulation data and extract 
physics reliably, and that these methods produce results in agreement
with well-established results.  It is for the last reason
that this article focuses on the $B_s$ system on quenched lattices, where 
methods existing in the literature provide a solid basis for
comparison.   We present results for $B_s$ meson kinetic 
masses, some level splittings, and the decay constant $f_{B_s}$ 
as evidence that our approach is working.  For quick reference,
we summarize the results of our finest isotropic lattice
in Table~\ref{tab:summary1232}.
Note that several systematic uncertainties remain, notably
the error from determining the spacing on quenched lattices
and discretization errors from using coarse lattices with the
present level of improvement.
Therefore, the results presented in this paper are useful for comparison
to similar lattice calculations, but they are not appropriate for inclusion
in phenomenological analyses.  Having established this method as a 
promising approach, work is now underway on unquenched lattices
to remove or reduce the systematic uncertainties and obtain 
state-of-the-art lattice QCD results.

In the next section we introduce and describe the formalism for 
combining staggered light quarks with heavy quark fields to form 
bilinear operators that create heavy-light mesons or represent 
heavy-light currents.   A significant simplification comes about 
from recognizing the equivalence of staggered and naive fermions and
 writing down bilinears in terms of the latter.  This will be explained 
thoroughly below.  In Section \ref{sec:simul} we give simulation 
details starting 
with a description of the glue, heavy quark, and light quark actions 
and then a discussion of our constrained fitting methods based on 
Bayesian statistics.  Section \ref{sec:results} gives results for 
heavy-light spectroscopy, including kinetic masses
and a calculation of the $B_s$ meson decay constant $f_{B_s}$.  
Three appendices contain details regarding the theory, notation,
and fitting techniques, respectively.


\section{Formalism}
\label{sec:formalism}

In this section we describe how to combine naive/staggered light quarks with 
heavy quarks to form heavy-light meson and electroweak current operators. 
We adhere to the recently introduced practice of calling the doubler 
degrees of freedom 
 ``tastes'' rather than ``flavors'' \cite{BernardLepage:2002}
(see also \cite{Aubin:2002ss}).
  We will be guided by 
the following properties of naive/staggered actions.

\begin{enumerate}
\item  Up to overall normalization factors, there is no difference between
using naive or staggered valence quarks in meson creation or 
current operators.  Since naive fermions are easier to interpret and to 
handle theoretically, we will 
construct our heavy-light bilinears using naive fermion fields rather 
than staggered fields.

\item  Any correlator involving naive fermion propagators can be 
rewritten in terms of staggered propagators.  Since staggered propagators 
are cheaper to calculate numerically, when it comes to actual simulations 
we will always 
work with expressions that have been converted to the staggered 
fermion language and involve only 
staggered (and heavy) quark propagators.  

\item  The taste content of naive/staggered actions can be 
determined either in the coordinate or the momentum space basis.  For 
heavy-light physics and for perturbation theory we find the 
momentum space interpretation to be more useful.  

\end{enumerate}

\noindent
We start by reviewing naive fermions 
and the identification of different tastes in momentum 
space. We will then
 explore the taste content of $B$ mesons 
that appear when naive fermions are combined with heavy fermions.
 We assume that the heavy quark action has no doublers, as in 
NRQCD, or that doublers have been given masses of order the cutoff 
via a Wilson term, as in the Fermilab approach \cite{El-Khadra:1997mp}.
Heavy-light systems
are much simpler than light-light systems since the heavy quark
suppresses the taste-changing processes of the naive/staggered quark.

\subsection{The free naive quark action}
Most of our discussion in this section will be for free unimproved 
naive fermions.  Taste identification and relevant symmetries 
survive the inclusion of gauge interactions and of the $\order{a^2}$ 
improvement terms incorporated into the action that we actually use in our 
simulations (see next section for a description of the full action).
The free unimproved naive fermion action is given by,

\beq
\label{nveaction}
{\cal S}_0 = a^4 \sum_x \left\{ \psiud (x) \,\left[ \sum_\mu \gamma_\mu 
{\frac{1}{a}}\nabla_\mu \; + \; m \right] \Psi(x) \right\},
\eeq
with
\beq
\nabla_\mu \Psi(x) = \half \,[ \Psi(x+a_\mu) - \Psi(x-a_\mu)].
\eeq
We work with hermitian Euclidean $\gamma$-matrices obeying 
$\{\gamma_\mu,\gamma_\nu \} = 2 \delta_{\mu \nu}$. 
It is well-known that the action (\ref{nveaction}) 
describes a theory with 16 tastes of 
Dirac fermions and that it has a set of discrete ``doubling'' 
symmetries,
\bea
\label{dsymm}
\Psi(x) & \rightarrow & e^{ix\cdot\pi_g}M_g \Psi(x)  \nl
\psiud(x) & \rightarrow & e^{ix\cdot\pi_g} \psiud(x) M_g^\dagger \,.
\eea
$g$ is an element of $G$, the set of ordered 
lists of up to 4 indices,
\beq
\label{setg}
G = \left\{ g: g = (\mu_1, \mu_2, \ldots ),\; \mu_1 < \mu_2 < 
\ldots \right\} \; ;
\eeq
e.g.\ $(2)$, $(0,3)$, and $(0,1,2,3)$ are elements of $G$, as is
the empty set $\emptyset$.
The corners of the Brillouin zone are denoted by
the 4-vector $\pi_g$ such that
\beq
\left(\pi_g\right)_\mu = 
\cases{
\frac{\pi}{a} \qquad \mu \in g, \cr
0 \qquad {\rm otherwise}.  \cr }
\eeq
The $M_g$ are transformation matrices
\beq
\label{mmat}
M_g = \prod_{\mu\in g} ~M_{\mu}
\eeq
with
\beq
M_\mu = i \gamma_5 \gamma_\mu .
\eeq

An illustrative way to reduce the taste degeneracy of the
naive action is to diagonalize the action in spin space.
Let $\phiud(x)$ and $\Phi(x)$ be a new set of 4-component 
spinor fields related to the original $\psiud(x)$ and $\Psi(x)$ 
fields via the Kawamoto-Smit \cite{Kawamoto:1981hw} transformation.
\beq
\label{spindiag}
\Psi(x) = \Omega(x) \, \Phi(x)   \qquad \qquad
\psiud(x) = \phiud(x) \, \Omega(x)^\dagger
\eeq
with
\beq
\label{omeg}
\Omega(x) = \prod^3_{\mu=0 } (\gamma_\mu)^{x_\mu/a} \, .
\eeq
In terms of these new fields the naive fermion action takes on a 
spin-diagonal form,
\beq
\label{phiaction}
{\cal S}_0 \rightarrow {\cal S}_\Phi = 
 a^4 \sum_x \left\{ \phiud (x) \,\left[ \sum_\mu \eta_\mu(x) 
{\frac{1}{a}}\nabla_\mu \; + \; m \right] \Phi(x) \right\},
\eeq
where
\beq
\eta_\mu(x) = (-1)^{(x_0 + \ldots + x_{\mu-1})/a} \, .
\eeq

Staggered fermions reduce the taste-degeneracy from 16-fold to
4-fold.  The spin-diagonal form of Eq.~(\ref{phiaction}) tells us 
it should be possible to do so,  since each spin component of 
$\Phi(x)$ is independent of the other components.  
One way to proceed is to define 1-component fields $\chi(x)$ through
\beq
\Phi(x) \equiv e(x) \chi(x) \, .
\eeq
The c-number spinor $e(x)$  is usually chosen to be constant,
 and one ends up with the standard staggered fermion action 
for the fields $\chi(x)$.  Reference \cite{Sharatchandra:1981si} 
goes through a more 
rigorous and general method for reducing the number of independent 
tastes from 16 to 4 which does not rely on first going through
the Kawamoto-Smit transformation.
They exploit the symmetry (\ref{dsymm}) to place 
constraints among the 16 different tastes so that only 4 of them remain 
as independent degrees of freedom.
(See also \cite{Chodos:1977yx} which uses the Hamiltonian formalism .)    

Equations (\ref{spindiag}) and (\ref{phiaction}) allow us 
to derive the simple but 
important relation between the naive propagator $G_\Psi(x;y)$
 and the staggered propagator $G_\chi(x;y)$. One has,
\bea
{\rm Eq.}~(\ref{spindiag}) \quad &\Longrightarrow& \quad
  G_\Psi(x;y) \; 
\equiv   \; \Omega(x) \; G_\Phi(x;y) \;  \Omega(y)^\dagger
  \\
{\rm Eq.}~(\ref{phiaction}) \quad &\Longrightarrow& \quad
 G_\Phi(x;y)  \; \equiv \; \hat{I}_4 \; G_\chi(x;y) \, ,
\eea
with $\hat{I}_4$ equal to a $4 \times 4$ identity matrix in Dirac space.
This leads to 
\beq
\label{nvestaggprop}
G_\Psi(x;y) \; \equiv \; \Omega(x) \Omega^\dagger(y) \;\times \;
 G_\chi(x;y).
\eeq
We use the identity (\ref{nvestaggprop}) repeatedly in the present work 
to go from bilinears expressed in terms of naive fermion fields to 
correlators written in terms of staggered propagators.  It can also be 
used to rederive familiar staggered correlators (e.g.\ for pions or 
rhos) starting from simple naive fermion bilinears.  We emphasize that 
(\ref{nvestaggprop}) is an exact relation even in the presence of gauge
 interactions; re-expressed as a relation between the inverse of the 
naive and staggered actions, respectively, for fixed  gauge fields, 
it is valid configuration by configuration, and hence also for the fully 
interacting naive and staggered propagators.
The relation (\ref{nvestaggprop}) also holds for improved versions of 
naive/staggered actions.

Before going on to discuss heavy-light bilinears, we end this subsection 
on basic naive fermion properties by reviewing the momentum space 
identification of naive fermion tastes. We continue to use the notation 
of \cite{Sharatchandra:1981si}.
The momentum space spinors are given by,
\beq
\label{psik}
\psi(k) = a^4 \sum_x e^{-ik\cdot x} \Psi(x) \qquad , \qquad
\psid(k) = a^4 \sum_x e^{ik\cdot x} \psiud(x)
\eeq
with the inverse relation given by
\beq
\Psi(x) = \int_{k,D} e^{ik \cdot x} \psi(k) \qquad , \qquad 
\psiud(x) = \int_{k,D} e^{-ik \cdot x} \psid(k).
\eeq
We  use the notation,
\beq 
\label{int4k}
\int_{k,D} \equiv \int _{k \in D} \frac{d^4k}{(2 \pi)^4} \qquad , \qquad
\int_{k,D_\emptyset} \equiv \int _{k \in D_\emptyset} \frac{d^4k}{(2 \pi)^4}
\eeq
where $D$ denotes the full Brillouin zone, 
$-\frac{\pi}{a} \le k_\mu < \frac{\pi}{a}$, and 
 $D_\emptyset$ just the central region, 
$-\frac{\pi}{2a} \le k_\mu < \frac{\pi}{2a} $.  
In terms of the momentum space spinors the free action (\ref{nveaction})
becomes
\beq
\label{momaction}
{\cal S}_0 = \int_{k,D} \psid(k) \left[ \sum_\mu i \gamma_\mu 
\frac{1}{a}\sin(k_\mu a) \; + \; m \right] \psi(k)
\eeq
Using the 4-vectors $\pi_g$ this can be written as,
\beq
\label{momact2}
{\cal S}_0 = \sum_g \int_{k,D_\emptyset} \psid(k+\pi_g) \left[ \sum_\mu i 
\gamma_\mu 
\frac{1}{a}\sin([k + \pi_g]_\mu a) \; + \; m \right] \psi(k+\pi_g)
\eeq
The next step is to define 16 new momentum space spinors $q^g(k)$ labeled 
by the elements $g$ of the set $G$ (\ref{setg})
\beq
\label{qspinors}
  q^g(k) = M_g \psi(k + \pi_g) \qquad , \qquad 
  \qd^g(k) = \psid(k + \pi_g) M_g^\dagger \, ;
\eeq
the matrices $M_g$ are those of (\ref{mmat}). 
 In terms of these new spinors, $q^g(k)$, and upon using the relation
\beq
M_g \gamma_\mu M_g^\dagger \, \sin([k + \pi_g]_\mu a) = \gamma_\mu
\, \sin(k_\mu a),
\eeq
the action ${\cal S}_0$ becomes
\beq
\label{momactq}
{\cal S}_0 =\sum_g \int_{k,D_\emptyset} \qd^g(k) 
\left[ \sum_\mu i \gamma_\mu 
\frac{1}{a}\sin(k_\mu a) \; + \; m \right] q^g(k).
\eeq
Eq.~(\ref{momactq}) clearly describes an action for 16 ``tastes''
of Dirac fermions.  The sum $\sum_g$ over the elements of the set $G$ 
can be interpreted as a sum over tastes.
The doubling symmetry (\ref{dsymm}) which in momentum space 
 becomes
\begin{eqnarray}
\label{dsymmp}
\psi(k) & \rightarrow & M_g \psi(k + \pi_g)  \nl
\psid(k) & \rightarrow &  \psid(k + \pi_g) M_g^\dagger,
\end{eqnarray}
takes one $q^g(k)$ taste into another up to possible sign factors, 
$\epsilon_{g_1,g_2} = \pm 1$, defined through $M_{g_1}M_{g_2} = \epsilon_{g_1,g_2} 
M_{g_1g_2}$ 
(see Ref.~\cite{Sharatchandra:1981si}).

\subsection{Heavy-light bilinears}

To discuss heavy-light bilinears we introduce heavy quark fields $\Psi_H$, 
which can stand for either heavy Wilson or nonrelativistic fermions 
(for the latter case 
we will use the notation $\Psi_H(x) \rightarrow Q(x)$ in later sections 
with $Q(x)$ a 4-component spinor with vanishing lower 2 components).
The simplest interpolating operator one could write down
 for creating a $B$ meson with 
a heavy quark field $\hpsid(x)$ and a naive antiquark field $\Psi(x)$ is
\beq
\label{bm}
{\cal W}_{B}(x) = \hpsid(x) \gamma_5  \Psi(x) \, .
\eeq

Let us analyze ${\cal W}_{B}(x)$ in 3-dimensional momentum space.
To do so we introduce the 3D Fourier transformed fields
\beq
\tilde{\psi}(\kv,t) = a^3 \sum_{\xv} e^{-i \kv \cdot \xv} \; \Psi(\xv,t) 
\qquad , \qquad \overline{\tilde{\psi}}(\kv,t) = a^3 \sum_{\xv}
e^{i \kv \cdot \xv } \; \psiud(\xv,t)
\label{eq:3dtilde}
\eeq
and similarly for the heavy fields $\Psi_H$.
It is useful to introduce a subset $G_s \subset G$ that involves only 
spatial indices $\mu \rightarrow  j=1,2,3$.  The full set $G$ can be built up 
 out of $g_s$ and $g_t g_s$ with $g_s \in G_s$ and  $g_t$ corresponding to 
$\mu = 0$ (and $M_{g_t} = i \gamma_5 \gamma_0$).  In analogy with
(\ref{int4k}) we have
\beq
\label{int3k}
\int_{\kv,D_s} \equiv \int _{\kv \in D_s} \frac{d^3k}{(2 \pi)^3}
 \qquad , \qquad
\int_{\kv,D_{s,\emptyset}} \equiv \int _{\kv \in D_{s,\emptyset}}
 \frac{d^3k}{(2 \pi)^3}
\eeq
where $D_s$ denotes the full 3D Brillouin zone, 
$-\frac{\pi}{a} \le k_j < \frac{\pi}{a}$, and 
 $D_{s,\emptyset}$  the central region, 
$-\frac{\pi}{2a} \le k_j < \frac{\pi}{2a}$.    
Then, as is shown in detail in Appendix~\ref{sec:formaldetail},
\begin{eqnarray}
\label{bgeq0}
a^3 \, \sum_{\xv} \wcal_{B}(\xv,t) & = &
 \sum_{g_s \in G_s} \int_{\kv,D_{s,\emptyset}} 
  \hktint e^{ik_0t} \nl
 & & 
 \left\{\overline{\tilde{\psi}}_H(\kv + \piv_{g_s},t) \,  \gamma_5
 \left[ M_{g_s}^\dagger q^{g_s}(\kv,k_0) + 
(-1)^t M^\dagger_{g_tg_s} q^{g_tg_s}(\kv,k_0) \right]   \right\} \, .
\end{eqnarray}
For $g_s\ne\emptyset$, the field 
$\overline{\tilde{\psi}}_H(\kv + \piv_{g_s},t)$ creates
a heavy quark with large spatial momentum 
so that any state containing it will have a large energy.
Consequently, the contributions to the heavy-light bilinear
$\sum_{\xv} \wcal_{B}(\xv,t)$ from 
low-lying states come from the $g_s=\emptyset$
part of the sum in (\ref{bgeq0}) 
\beq
\label{bgeq0gs0}
 \int_{\kv,D_{s,\emptyset}} 
  \hktint e^{ik_0t} 
 \left\{\overline{\tilde{\psi}}_H(\kv,t) \,  \gamma_5 \,
 \left[ q^{\emptyset}(\kv,k_0) + 
(-1)^t M^\dagger_{g_t} q^{g_t}(\kv,k_0) \right]   \right\} \,.
\eeq
In contrast, light-light bilinears receive contributions from
all 8 sections of the spatial Brillouin zone (this can be seen
by replacing $\overline{\tilde{\psi}}_H$ by $\overline{\tilde{\psi}}$
in (\ref{bgeq0}) and then using (\ref{qspinors})).
The $g_s\ne 0$ contributions to heavy-light bilinears are discussed in
more detail in Appendix~\ref{sec:formaldetail}, where
we consider more general
bilinears and show that they couple either to exactly degenerate
states or to artificial high energy lattice states.

Let us point out that in (\ref{bgeq0gs0}) there are contributions 
from both the pseudoscalar and the scalar state, which has a 
coefficient alternating in sign.  The oscillating parity partner 
appears in light-light correlators as well.  In Section~\ref{sec:fitting}
we discuss how fits are able to separate these contributions
from correlation functions.

\subsection{Heavy-light two-point correlators}

\noindent
Once heavy-light bilinears with naive light quarks have been 
introduced, it is straightforward to obtain 
 bilinear-bilinear two-point correlators and write them in terms of 
staggered propagators. Starting from this subsection we will 
revert to the usual practice of working with dimensionless spinor 
fields.  Hence one should assume all $\Psi$, $\Psi_H$ and $\chi$ fields 
have been multiplied by a factor of $a^{3/2}$ and that all 
propagators are now dimensionless. 
Denoting the generic bilinear as
\beq 
{\cal W}_\Gamma(x) = \hpsid(x) \Gamma  \Psi(x),
\eeq
one has
\bea
\sum_{\xv} e^{i \pv \cdot \xv} 
\langle {\cal W}^\dagger_{\Gamma_{sk}}(x) \;
{\cal W}_{\Gamma_{sc}}(0) \rangle  & = & 
\sum_{\xv} e^{i \pv \cdot \xv} {\rm Tr} \left\{\Gamma_{sc} \,  G_\Psi(0;x) 
\, \Gamma_{sk}^\dagger \, G_H(x;0) \right\}  \nl
& =& 
\sum_{\xv} e^{i \pv \cdot \xv} 
\sum_{c,c^\prime} \left [ {\rm tr} \left\{ \Gamma_{sc} \, \Omega^\dagger(x) \,
\Gamma_{sk}^\dagger \, G_H^{c^\prime c}(x;0) \right\} \; 
G_\chi^{cc^\prime}(0;x) \right ],
\eea
where we have used eq.(\ref{nvestaggprop}) to convert from $G_\Psi$ 
to $G_\chi$.  ``Tr'' stands for a trace over both color and spin indices,
whereas ``tr'' stands for a trace over spin indices only.
Using $G_\chi(0;x) = G^\dagger_\chi(x;0) (-1)^{\sum_\mu x_\mu/a}$ 
one gets for the case $\Gamma_{sc} = \Gamma_{sk} = \gamma_5$
\bea
C^{(2)}_{B}(\pv,t) & = & 
\sum_{\xv} e^{i \pv \cdot \xv} 
\langle {\cal W}^\dagger_B(x) \;
{\cal W}_B(0) \rangle  \nl
& = & 
\sum_{\xv} e^{i \pv \cdot \xv} 
\sum_{c,c^\prime} \left [ {\rm tr} \left\{  \Omega^\dagger(x) 
 \, G_H^{c^\prime c}(x;0) \right\} \; 
G_\chi^{*c^\prime c}(x;0) \right ] 
\eea
which couples to the $B$ meson. For the $B^*$ meson, we set
$\Gamma_{sc} = \Gamma_{sk} = \gamma_j$ which gives
\beq
C^{(2)}_{B^*}(\pv,t)  =  
\sum_{\xv} e^{i \pv \cdot \xv} 
\sum_{c,c^\prime} \left [ {\rm tr} \left\{  \Omega^\dagger(x) 
 \, G_H^{c^\prime c}(x;0) \right\} \; (-1)^{x_j/a} \;
G_\chi^{*c^\prime c}(x;0) \right ] .
\eeq

In the above formulas we are now allowing the heavy-light mesons 
to have nontrivial momentum. 
 As long as spatial momenta 
are restricted to $ap_j < \pi/2$ there should be no problems with 
 the Lorentz and/or taste content of a meson suddenly changing at 
finite momenta. In later sections of this article we will present 
results showing good dispersion relations for $B$ and $B^*$ mesons 
for momenta up to at least $ap_j = \pi/3$ to check this 
hypothesis.

Although the discussion above implicitly assumes the use of
local sources and sinks, generalizing to smeared sources and
sinks is straightforward as long as one takes care that the
smearing function preserves the doubling symmetry (\ref{dsymm}).
This work employs local sources and sinks, with good results for
the ground state mesons, but smearing is an important direction
for future studies, especially those of excited states. 
Nonlocal sources have been used extensively in staggered fermion
simulations of light hadrons.


\section{Simulation details}
\label{sec:simul}

\subsection{Actions and parameters}

The gauge action used to generate the isotropic gauge configurations
is the tadpole-improved tree-level $\order{a^2}$-improved action
\cite{Weisz:1983zw,Weisz:1984bn}
\begin{equation}
\act^{(iso)}_G ~=~ - \beta \sum_{x,\mu > \nu}  \left\{
\frac{5}{3} \frac{P_{\mu\nu}(x)}{u_\mu^2 u_\nu^2} 
- \frac{1}{12} \frac{R_{\mu\nu}(x)}{u_\mu^4 u_\nu^2} 
- \frac{1}{12} \frac{R_{\nu\mu}(x)}{u_\nu^4 u_\mu^2} \right\} \, .
\end{equation}
$P_{\mu\nu}$ represents the plaquette and $R_{\mu\nu}$ 
the $2\times 1$ rectangle in the $(\mu,\nu)$ plane; both
are normalized so that $\vev{P_{\mu\nu}} = \vev{R_{\mu\nu}} =1$ 
in the $\beta\to\infty$ limit.  As part of our tests 
we also study anisotropic lattices
where the temporal lattice spacing $a_t$ is a few times smaller
than the spatial lattice spacing $a_s$; in this case improvement in the
temporal direction is secondary to spatial improvement.
The action used for the anisotropic lattices is the same in the 
spatial directions, but the rectangles with two units in the
temporal direction are omitted and the space-time coefficients
adjusted to be consistent with Symanzik improvement 
\cite{Morningstar:1997ze,Alford:1997nx}:
\begin{eqnarray}
\act^{(aniso)}_G &=& - \beta \sum_{x,s > s^\prime} \frac{1}{\xi_0} \left\{
\frac{5}{3} \frac{P_{ss^\prime}(x)}{u_s^4} 
- \frac{1}{12} \frac{R_{ss^\prime}(x)}{u_s^6} 
- \frac{1}{12} \frac{R_{s^\prime s}(x)}{u_s^6} \right\} \nl
 & & - \beta \;\; \sum_{x,s} \xi_0 \left\{
\frac{4}{3} \frac{P_{st}(x)}{u_s^2 u_t^2} 
- \frac{1}{12} \frac{R_{st}(x)}{u_s^4 u_t^2} \right\} \, .
\end{eqnarray}
For the values of the inverse coupling $\beta$ and the bare
anisotropy $\xi_0$ used in this work, the tadpole-improvement
Landau-link factors $u_s$ and $u_t$, 
the spatial lattice spacing $a_s$, and the renormalized 
anisotropy $\xi \equiv a_s/a_t$ were determined in 
Ref.\ \cite{Alford:2000an}.
The simulation parameters for the gauge configurations are summarized
in Table~\ref{tab:runlist}.

The parameters for the isotropic lattices were intended to give
approximately the same spatial lattice spacings as the anisotropic
lattices.  The isotropic $8^3\times 20$ lattice parameters were discussed
in Ref.\ \cite{Alford:1998yy}.
The isotropic $12^3\times 32$ configurations were generated for
this work, and we
determined the lattice spacing by calculating the static quark potential
and using the phenomenological parameter $r_0=0.5$ fm \cite{Sommer:1994ce}
to set the scale.

The light quark action we use is the $\order{a^2}$ tadpole-improved
staggered action \cite{Orginos:1999cr,Lepage:1998vj} which contains
in place of 
the simple covariant difference operator in (\ref{nveaction}) 
an improved difference operator constructed as follows.  First, the 
link matrices $U_\mu(x)$
are replaced by ``fat-link'' matrices \cite{Blum:1997uf}:
\beq
V_\mu(x) ~\equiv~ \prod_{\rho\ne\mu} \left(1 + \frac{\nabla^{(2)}_\rho}{4}
\right)\bigg|_{\rm symmetrized} U_\mu(x)
\label{eq:fatlink}
\eeq
which contain 3, 5, and 7-link paths, all bent to fit within an
elemental hypercube (Ref.\ \cite{Orginos:1999cr} lists each term
explicitly, and we write the second-derivative operator $\nabla^{(2)}$
in Appendix~\ref{sec:derivatives}).  
This smearing effectively introduces a form factor
in the quark-gluon vertex which suppresses the coupling of high
momentum gluons to low momentum quarks.
Second, the fat-link is further modified by adding what has come to be known 
as the Lepage term \cite{Lepage:1998vj} in order to cancel 
the low momentum $\order{a^2}$ error introduced by (\ref{eq:fatlink}):
\beq
V_\mu'(x) ~\equiv~ V_\mu(x) - \sum_{\rho\ne\mu} \frac{(\nabla_\rho)^2}
{4} U_\mu(x) \, .
\eeq
Finally, the remaining $\order{a^2}$ (rotational) 
errors are subtracted by including a cube of the difference operator, the
so-called Naik term \cite{Naik:1989bn}; therefore the $\order{a^2}$
improved action is obtained by the replacement
\beq
\nabla_\mu ~\longrightarrow~ \nabla_\mu' - \frac{1}{6}(\nabla_\mu)^3 \,.
\label{eq:naik}
\eeq
This action has been used in many recent simulations, 
quenched and unquenched, most prominently by the MILC Collaboration
who call it the ``AsqTad'' action.
In order to apply tadpole improvement consistently,
powers of the covariant difference operators, $(\nabla_\mu)^n$
and $(\nabla_\mu^{(2)})^n$,
are obtained by $n$ successive applications of $\nabla_\mu$
or $\nabla_\mu^{(2)}$, respectively,
with no tadpole factors, replacing $U_\mu \to U_\mu/u_\mu$
in the final expression only after setting terms like
$U_\mu(x)U_\mu^\dagger(x)$ equal to 1.  In other words, one writes
every operator in (\ref{eq:fatlink}) in terms of paths of links, 
dividing each link variable by its corresponding tadpole factor $u_\mu$.

In this work we utilize anisotropic lattices, for which the improved
staggered action is rewritten breaking the sum over spacetime directions
into spatial and temporal parts
\beq
a_t\sum_\mu \frac{\eta_\mu}{a_\mu}
\left( \nabla_\mu' - \frac{1}{6}(\nabla_\mu)^3 \right)
~\longrightarrow~ 
\eta_t\left( \nabla_t' - \tnaik\frac{1}{6}(\nabla_t)^3 \right)
\;+\; \frac{c_0}{\xi} 
\sum_k \eta_k\left( \nabla_k' - \frac{1}{6}(\nabla_k)^3 \right)
\label{eq:anisoasqtad}
\eeq
The parameter $c_0$ is tuned to give the correct pion dispersion relation.
We include a parameter $\tnaik$ which we set equal to 1 or 0 whether we
want to include the 3-link hopping in the temporal direction or not;
we still call the $\tnaik=1$ action ``AsqTad'', and we refer to
the $\tnaik=0$ action as ``AsqTad-tn.''
Note that the  isotropic AsqTad
action is recovered by setting $c_0 = \xi = \tnaik = 1$.

The NRQCD action is \cite{Lepage:1992tx,Collins:2001pe}
\bea
 \label{nrqcdact}
\act_{\rm NRQCD}  = 
\sum_x \Bigg\{  {\phi}^\dagger_t \phi_t &-& 
 {\phi}^\dagger_t
\left(1 \!-\!\frac{a_t \delta H}{2}\right)_t
 \left(1\!-\!\frac{a_tH_0}{2n}\right)^{n}_t \nonumber \\
& \times &
 U^\dagger_t(t-1)
 \left(1\!-\!\frac{a_tH_0}{2n}\right)^{n}_{t-1}
\left(1\!-\!\frac{a_t\delta H}{2}\right)_{t-1} \phi_{t-1} \Bigg\} \, .
 \eea
 $H_0$ is the nonrelativistic kinetic energy operator,
 \beq
a_t H_0 = - {\delsq\over2\xi\Mbz}
 \eeq
and $\delta H$ includes relativistic and finite-lattice-spacing
corrections,
 \begin{eqnarray}
a_t\delta H 
&=& - c_1\,\frac{(\delsq)^2}{8\xi\Mbz^3} 
+ c_2\,\frac{i}{8\Mbz^2}\left(\delv\cdot\Ev - \Ev\cdot\delv\right) \nl
& &
 - c_3\,\frac{1}{8\Mbz^2} \sigmav\cdot(\delvt\times\Ev - \Ev\times\delvt)\nl
& & - c_4\,\frac{1}{2\xi\Mbz}\,\sigmav\cdot\Bv 
  + c_5\,\frac{\delfour}{24\xi\Mbz}  - c_6\,\frac{(\delsq)^2}
{16n\xi^2\Mbz^2} \, .
\label{deltaH}
\end{eqnarray}
All derivatives are tadpole improved and,
\beq
\delsq = \sum_{j=1}^3\nabla_j^{(2)}, \qquad \qquad \delfour = \sum_{j=1}^3
\nabla_j^{(4)}
\eeq
\beq
\tilde{\nabla}_k = \nabla_k - \frac{1}{6}\nabla_k^{(3)}
\eeq
The dimensionless Euclidean electric and magnetic fields are,
\beq
\tilde{E}_k = \tilde{F}_{k4}, \qquad \qquad \tilde{B}_k =
-\half \epsilon_{ijk}\tilde{F}_{ij} \;.
\eeq
Explicit expressions for 
$\nabla_k^{(m)},$ $m=2,3,4$ and $\tilde{F}_{\mu \nu}$ are 
given in Appendix~\ref{sec:derivatives}.  
In most cases we set all 6 of the $c_i = 1$ and refer to this 
as the $1/M^2$ NRQCD action, even though the leading $1/M^3$
relativistic correction is also included.  In order to make 
corresponding perturbative calculations simpler, some simulations
were done setting $c_1 = c_2 = c_3 = c_6 = 0$ with $c_4=c_5=1$,
and we call this the $1/M$ NRQCD action.  In practice the
results depend very little on which action is used, since the
nonrelativistic approximation is very good for $B$ mesons.

The bare mass of the heavy quark, $M_0$, is chosen to be close
to the bottom quark mass, based on simulations with Wilson-like
light quarks \cite{Collins:2001pe,Shigemitsu:2002wh}.  The bare 
mass of the staggered quark $m$ is tuned to be close to
the strange quark mass using the condition that the ratio of the
``$\bar{s}s$'' pseudoscalar meson mass to the ``$\bar{s}s$''
vector meson mass is approximately equal to
$\sqrt{2m_K^2 - m_\pi^2}/m_\phi = 0.673$.  On unquenched lattices
the $\phi$ mass is probably not accurately determined since 
it should be sensitive to the sea quark masses decreasing through
the threshold for $\phi\to K K$.  Instead one should first determine
the lattice spacing, then use $2m_K^2 - m_\pi^2$ to 
determine the bare strange quark mass.  On the other hand,
for the quenched lattices
in this work, the ratio serves as an appropriate
fiducial for comparison between different lattices.

\subsection{Fitting methods}
\label{sec:fitting}

The light quark propagators are computed with anti-periodic boundary 
conditions in the temporal direction; in contrast, the 
evolution of the heavy quark in time requires only an initial
condition.  Due to this difference, heavy-light meson correlators
with temporal separations greater than $N_t/2$ will be 
contaminated from the light quark propagating backward in time
from the source across the time boundary, so we only compute heavy-light
meson correlators up to $N_t/2$.  The periodicity of the light quark
can still be utilized to improve statistics
by evolving the heavy quark backward in time
from the source.  We average the forward and backward propagating
meson correlators configuration by configuration.

The process of fitting the meson correlators to a series of 
exponentials is complicated because
the temporal doubler causes the correlation function to couple
not only to states with the quantum numbers expected from
the continuum limit, but also to states with opposite parity
times an oscillating factor $(-1)^{t+1}$.  Thus,
we expect the meson correlators to have the form 
\beq
f(t;\{A_k,E_k\}) ~=~ \sum_{n=0}^{K_p-1} A_k e^{-E_k t} ~+~
\sum_{k=K_p}^{K_p + K_o -1}(-1)^{t+1} A_k e^{-E_k t} 
\label{eq:corrfnsum}
\eeq
which includes $K_p$ states of expected parity and $K_o$
states of opposite parity.  In our study we always take
$K_o = K_p$ or $K_p-1$,  and for the excited state energies we
use the differences $\Delta E_k \equiv E_k - E_{k-2}$ as
parameters in the fit.
The $K = K_p + K_o$ terms in the fitting function 
(\ref{eq:corrfnsum}) can be rearranged as
\bea
f(t;\{A_k,E_k\}) & = & {A_0} e^{-{E_0} t} 
~+~ (-1)^{t+1} {A_1} e^{-{E_1} t} \nonumber \\
&+& \sum_{k=2}^{K-1} (-1)^{k(t+1)} \, {A_k} 
\, e^{-(\Delta E_k + \Delta E_{k-2} + \ldots) \,t}  \, .
\label{eq:fitform}
\eea
Note that terms with even $k$ are simple exponentials and
those with odd $k$ are oscillating exponentials.

Recently a curve fitting method has been introduced to
our community which allows one to estimate the systematic
uncertainty from the series of states (\ref{eq:fitform})
in the correlator \cite{Lepage:2001ym,Morningstar:2001je}.
One fits the correlation function $C(\VEC{p},t)$ for
all computed values of $t$, varying the number of terms
$K$ in the fit.  For a given $K$, the best fit is obtained
by minimizing an augmented $\chi^2$:
\beq
\label{eq:chisqaug}
\chisqaug(C(t),\{\lambda_i\},\{(\mu_i,\delta_i)\}) 
~\equiv~ \chi^2(C(t),\{\lambda_i\}) ~+~ \sum_{i=0}^{2K-1} 
\frac{(\lambda_i - \mu_{\lambda_i})^2}{\delta_{\lambda_i}^2}
\; .
\eeq
where we have generically denoted the parameters of 
(\ref{eq:fitform}) by 
\beq
{\bm \lambda} \equiv (A_0, E_0, A_1, E_1, A_2, \Delta E_2, \ldots, 
A_K, \Delta E_K)
\label{eq:paramvector}
\eeq
the $i$-th component of which is $\lambda_i$.  

\begin{figure}
\begin{center}
\epsfxsize=\hsize
\mbox{\epsfbox{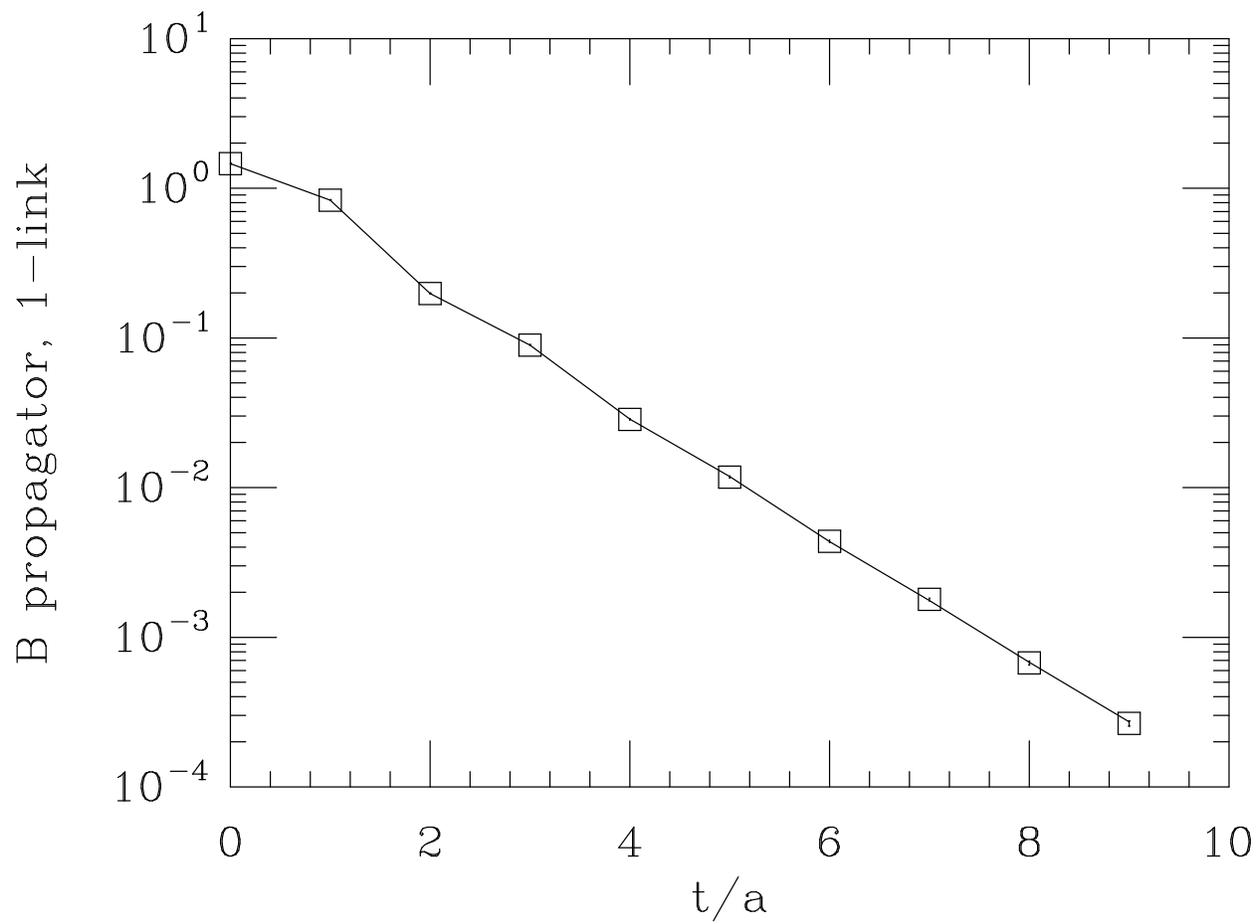}}
\caption{$B$ meson propagator on the \ainv = 0.8 GeV lattice
with a $am=0.18$ 1-link staggered quark and $aM_0=6.5$ nonrelativistic
heavy quark. The 3-exponential fit has $\chisqaugdof = 0.59$.}
\label{fig:bmesprop1l820}
\end{center}
\end{figure}

In Appendix~\ref{sec:fittingdetail} we give a pedagogical
summary of \cite{Lepage:2001ym} as it applies to our calculation, but
a few remarks here are in order.
The second term on the right-hand side of (\ref{eq:chisqaug})
is the contribution of Gaussian priors for each fit parameter,
and one sets the prior means $\mu_{\lambda_i}$ and half-widths
$\delta_{\lambda_i}$ based on reasonable prior estimates for
those quantities.  The procedure is best illustrated by an example.
Let us take a pseudoscalar heavy-light correlator, computed
with the unimproved, or ``1-link'' staggered action, on the 
$8^3\times 20$ lattice as an example (see Fig.\ \ref{fig:bmesprop1l820}).
The set of prior means $\mu$ and half-widths $\delta$ used in
fitting this correlator
is given in Table~\ref{tab:priors}.  The ground state energy
and amplitude prior means were estimated from effective mass plots
and the prior widths set at 50\% and 25\%, respectively.  Priors
for the excited states biased the amplitude fit parameters to
be all of the same order and the energy differences to be 
equal and about 300 MeV, roughly the size of the $2S-1S$ and
$1P-1S$ splittings in $B$ spectrum.  Recall that
the NRQCD action does not include the rest mass, so the energy
$E_{\rm sim}$ is equal to the physical meson mass minus an
energy shift $\Delta$.  
Tables \ref{tab:efitvsk}--\ref{tab:chisqaug}
show the results of fits to the propagator in Fig.\ \ref{fig:bmesprop1l820}
as the number of exponentials changes from 2 to 8.
The uncertainties
are estimated from the inverse of the matrix $\bm{\nabla \nabla} \chisqaug$
of second derivatives ($[{\bm{\nabla\nabla}} \chisqaug]_{ij} \equiv
\partial^2\chisqaug/\partial\lambda_i\partial\lambda_j$)
\beq
\sigma_{\lambda_i} ~=~ \sqrt{ 2
\left[\left(\bm{\nabla\nabla} \chisqaug\right)^{-1}\right]_{ii}} \;.
\label{eq:chiaug2deriv}
\eeq
which assumes the shape of $\chisqaug$ near its minimum 
$(\lambda_i=\lambda_i^{\rm min}, \,\forall_i)$ is
quadratic in $\lambda_i$
\beq
\chisqaug - \chisqaug|_{\rm min} ~\approx~ \frac{1}{2}\;\sum_{ij}\,
(\lambda_i - \lambda_i^{\rm min})\,
\frac{\partial^2\chisqaug}{\partial\lambda_i\partial\lambda_j}\,
(\lambda_j - \lambda_j^{\rm min}) \, .
\eeq

\begin{figure}
\begin{center}
\epsfxsize=\hsize
\mbox{\epsfbox{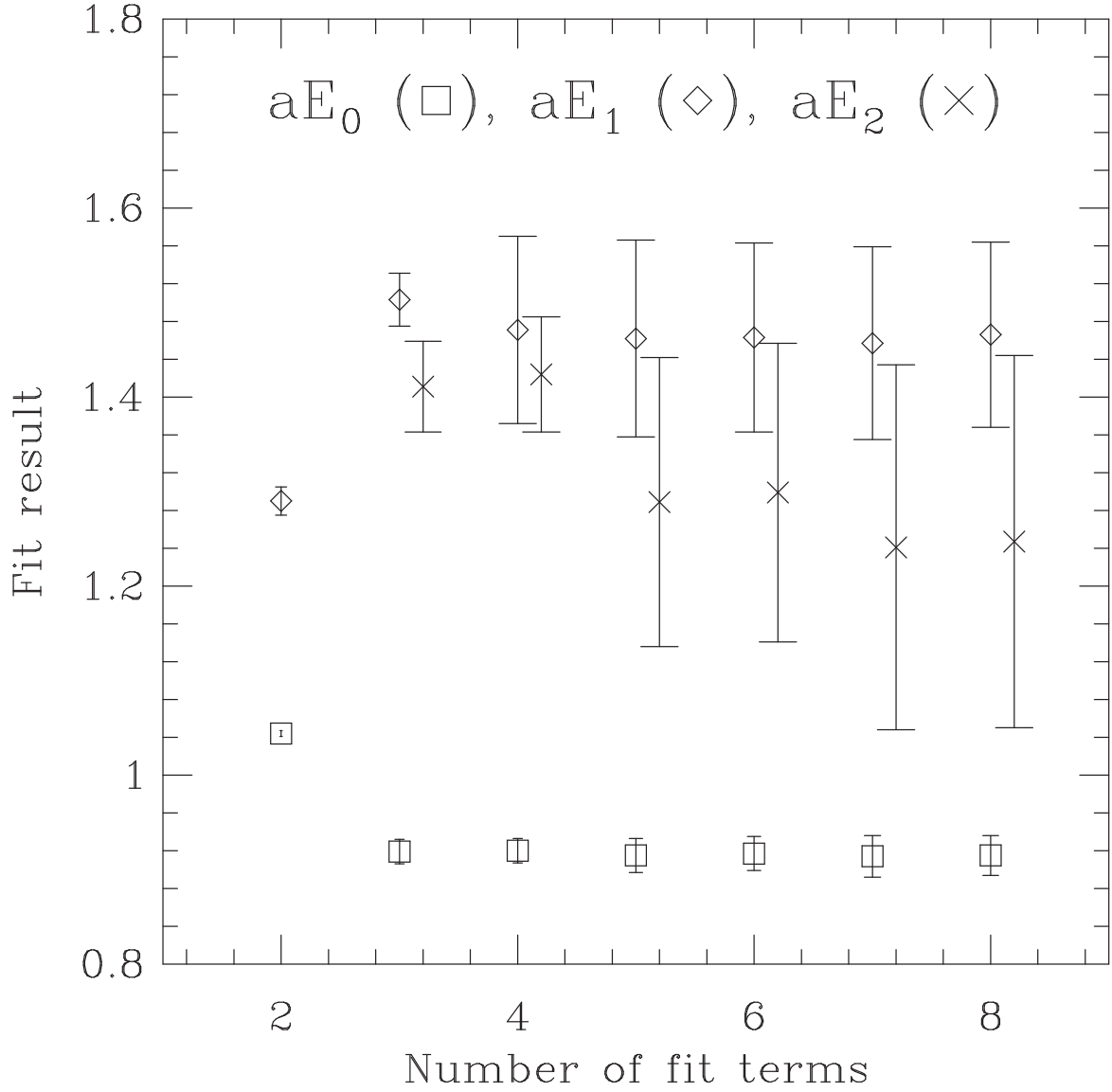}}
\caption{Results of several fits
to Propagator of Fig.\ \ref{fig:bmesprop1l820}
 plotted vs.\ number of exponentials in the fit. 
Parameters shown are the non-oscillating ground state energy
($aE_0$), the oscillating ground state energy ($aE_1$),
and the non-oscillating first excited state energy 
($aE_2 = aE_0 + a\Delta E_2$). }
\label{fig:fitvsK}
\end{center}
\end{figure}

In Fig.~\ref{fig:fitvsK} we plot the non-oscillating and oscillating
ground state energies, as well as the first excited state energy,
vs.\ the number $K$ of exponentials in the fit.  The rest of
the fit parameters are given in 
Tables \ref{tab:efitvsk} and \ref{tab:afitvsk}.
One can clearly see the stability of the ground state fit parameters
$A_0, E_0$ and $A_1, E_1$ as $K$ is increased.  The beginning
of a plateau at $K=3$ implies at least one excited
state is needed in the fit in order for the excited state effects
to be removed from the ground states.  Table \ref{tab:chisqaug}
similarly indicates that $K\ge 3$ is necessary in order to
have an ``acceptable'' $\chisqaugdof$; as we discuss in 
Appendix~\ref{sec:fittingdetail}, $\chisqaugdof$ should only be
used as a gross check of the fit.  E.g.\ $\chisqaugdof \agt 2$ 
implies the fit function is a highly improbable model of the
data, but one should not necessarily prefer a fit with 
$\chisqaugdof = 0.8$ over one with $\chisqaugdof = 1.3$.

Note that the uncertainties estimated from
the fit for the ground state parameters are much smaller than the
widths of the corresponding priors, while the errors from the fit
are comparable to the prior widths for most of the excited state
parameters.  The first excited non-oscillating state, $k=2$,
is an exception, appearing to be well constrained by the data until 
another non-oscillating state, $k=4$, is included in the fit.
This means that the $K=3$ and $K=4$ fit result for $E_2$ 
does a good job of absorbing the effects of the excited states, but
that there is not enough constraint from the data (or the priors)
to separate the first excited state from the second.  Thus,
we conclude that $K=3$ is sufficient to obtain reliable estimates
of the ground state energies and amplitudes and that the data is
not sufficiently precise to extract excited state energies and
amplitudes.

\begin{figure}
\begin{center}
\epsfxsize=\hsize
\mbox{\epsfbox{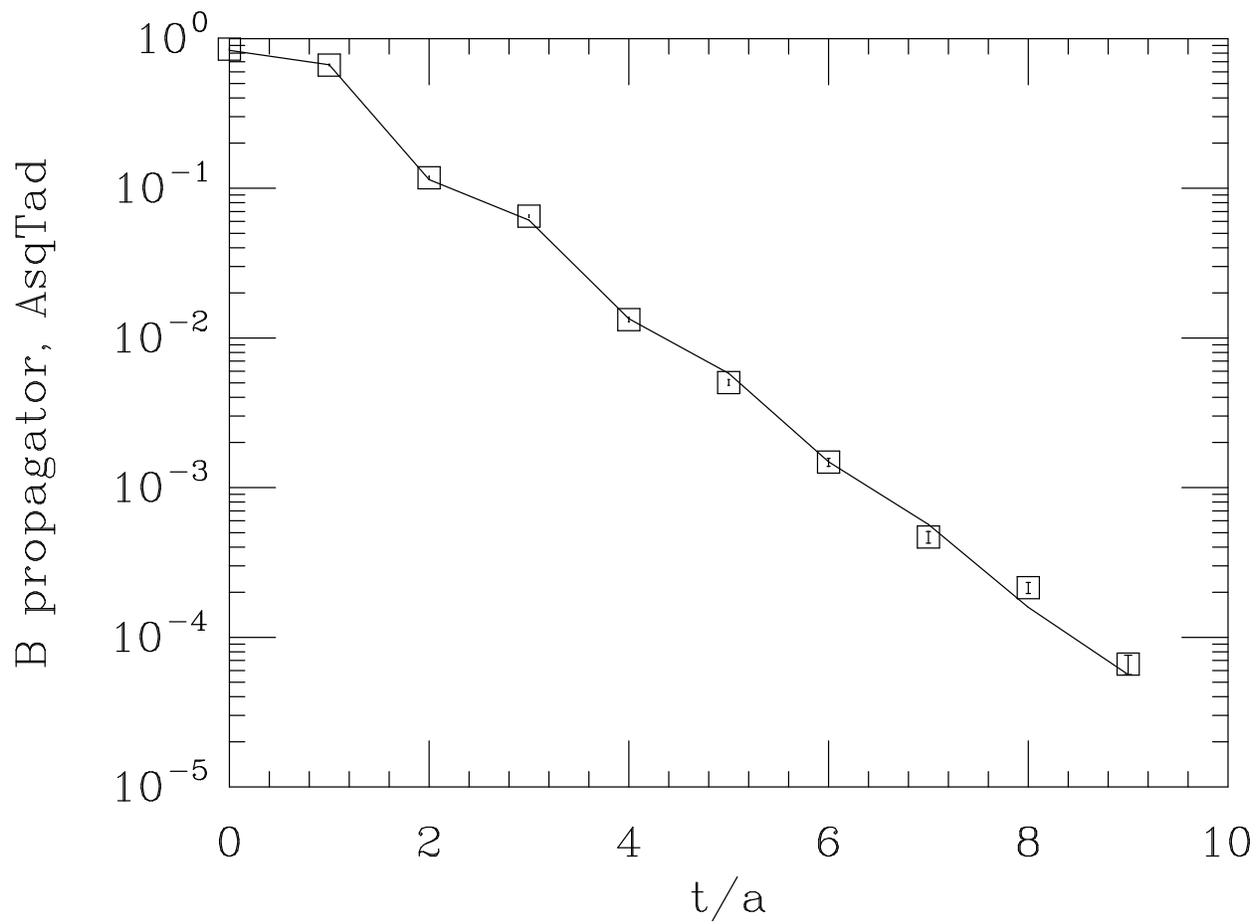}}
\caption{Example of the poor fits obtained for the 
$B$ meson propagator on the \ainv = 0.8 GeV lattice
with an improved staggered quark, caused by the temporal
Naik term on such coarse temporal lattice spacing (see text).
The fit shown has $\chisqaugdof = 8.9$.}
\label{fig:bmespropasq820}
\end{center}
\end{figure}

We are able to utilize this constrained curve fitting method
to fit all of our data except in one case: the heavy-light
correlators computed with the AsqTad action on the 
$8^3 \times 20$ lattice.  We were not able to find fits
with $\chisqaugdof < 8$; one example is shown
in Fig.\ \ref{fig:bmespropasq820} where the fit is visibly 
much worse than for the 1-link action shown 
in Fig.\ \ref{fig:bmesprop1l820}.  
This turns out to be a consequence of using
an action with next-to-next-to-nearest neighbor couplings in the
$\hat{t}$-direction on a lattice 
with coarse temporal lattice spacing.

\begin{figure}
\begin{center}
\epsfxsize=\hsize
\mbox{\epsfbox{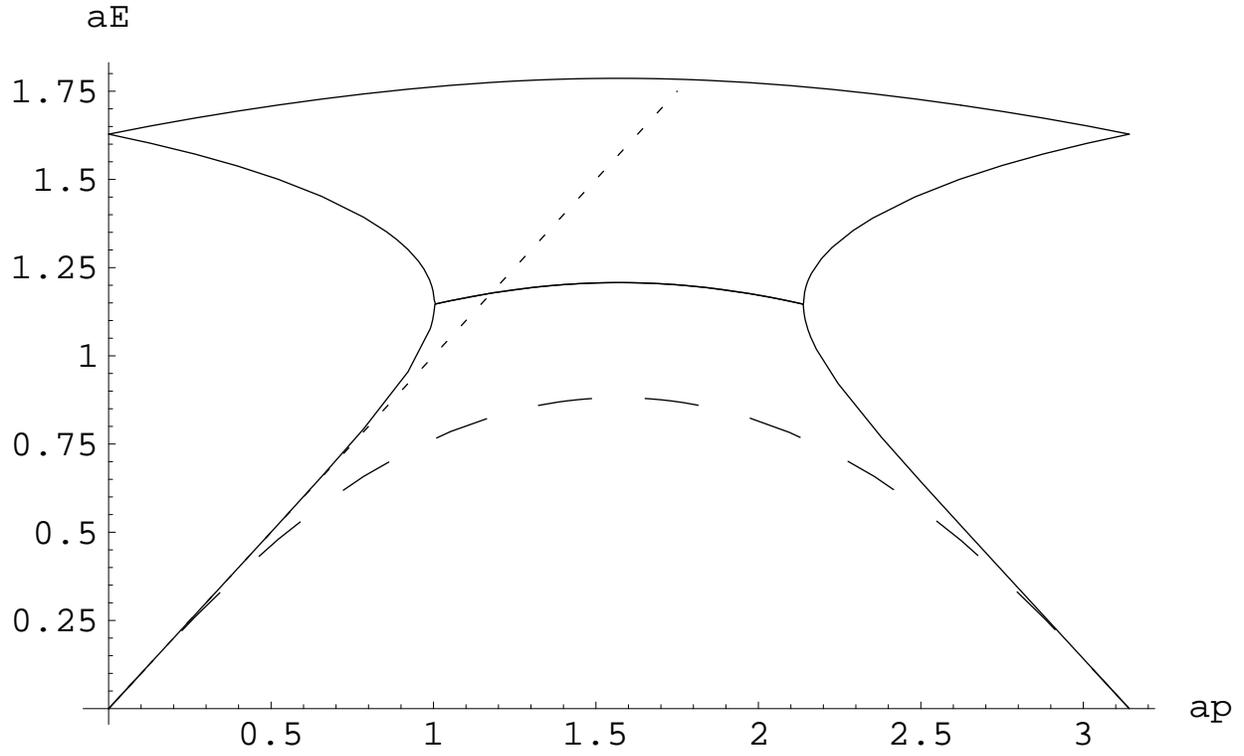}}
\caption{Dispersion relation for free massless fermions. The dotted
line shows the continuum dispersion relation $E^2=p^2$ , the 
dashed line shows the dispersion relation for the naive fermions
$\mathrm{sinh}^2 aE = \sin^2 ap$, and the solid lines show the real part
of the dispersion relation for the Naik action.  Note that the
solution of the Naik dispersion relation which most closely follows
the continuum dispersion relation is purely real until the branch point
near $ap\approx 1$.  }
\label{fig:naik_dispersion}
\end{center}
\end{figure}

The free Naik fermion dispersion relation 
(see Fig.\ \ref{fig:naik_dispersion}) has complex solutions
which implies there may be excited states with negative norms
contaminating the correlators at short time separations.  If the
temporal extent of the lattice were sufficiently long and sufficiently
precise correlators were computed, these negative norm states which
have energies proportional to $1/a$ would have a negligible effect:
one could include only points with $t$ greater than some minimum value
in the fit, or one could include a negative norm state in the fit.
However, for the $8^3\times 20$ lattice where $1/a=0.8$ GeV
we are unable to drop
enough points and get a good fit while keeping enough to fit to
states of both parities.  Also, when we tried to include a negative
norm exponential in the fit, large cancellations with the positive
norm excited states resulted in unstable fit results.

\begin{figure}
\begin{center}
\epsfxsize=\hsize
\mbox{\epsfbox{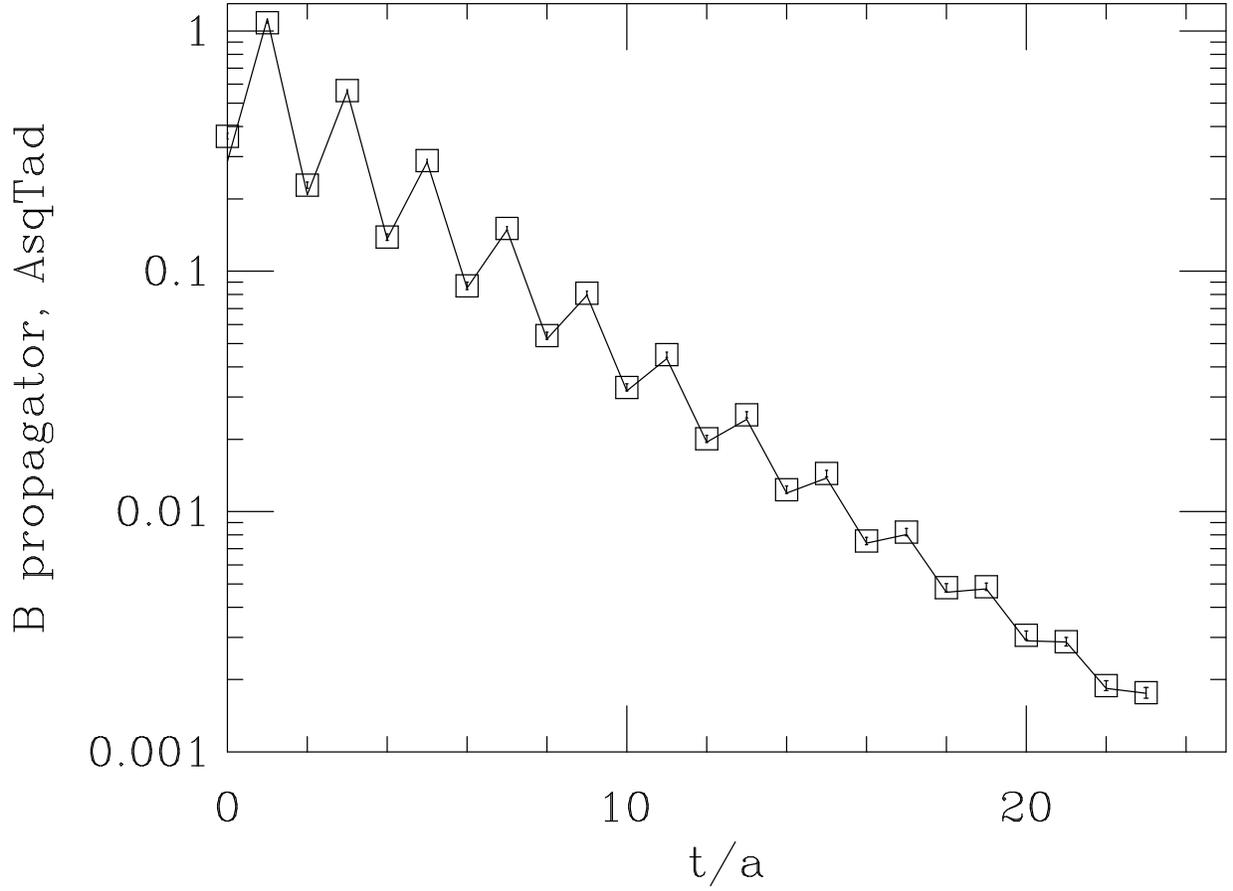}}
\caption{$B$ meson propagator on the anisotropic $8^3\times 48$ lattice,
where $a_t^{-1} = 3.7$ GeV,
with a $am=0.04$ improved staggered quark and $M_0=7.0$ nonrelativistic
heavy quark.  The lattice artifacts due to the temporal Naik term
do not contaminate the fit.  The 4-exponential fit plotted has
$\chisqaugdof = 0.87$.}
\label{fig:bmespropasq848}
\end{center}
\end{figure}

We checked this hypothesis on the $1/a=0.8$ GeV lattice
by simulating with 4 different staggered quark actions: 
the 1-link and improved actions as well as an action where the
Naik term was included but no fattening of the links was done
(the Naik action) and an action with fat-links but no Naik term
(the ``fat-link'' action).  We were able to obtain reasonable
fits to heavy-light correlators with the 1-link and fat-link
actions, but not with the Naik action nor the fully improved action;
we tabulate typical values for $\chisqaugdof$ in Table~\ref{tab:chisqs}.
Furthermore we performed simulations on an anisotropic $8^3\times 48$
lattice with a 
very fine temporal lattice spacing $1/a_t = 3.7$ GeV using 
the 1-link action, the AsqTad-tn action ($\tnaik = 0$),
and the full AsqTad action ($\tnaik = 1$).  
In all 3 cases we found acceptable
fits with similar values of $\chisqaugdof$, again tabulated in
Table~\ref{tab:chisqs}.
Fig.~\ref{fig:bmespropasq848} shows the pseudoscalar
propagator on this lattice for the AsqTad action.
We have no problem fitting to heavy-light correlators on a finer
isotropic $12^3\times 32$ lattice where $1/a = 1.0$ GeV
with the $\order{a^2}$ improved action.  Therefore, 
the origin of the problematic fits on the $1/a = 0.8$ GeV lattice
is due to a particular lattice artifact arising from the temporal Naik term;
but with a larger lattice scale $1/a \ge 1.0$ GeV
these artifacts become insignificant.

Let us return to the subject of estimating the uncertainties of the
fit parameters.  The second derivative
of $\chisqaug$ (\ref{eq:chiaug2deriv}) gives a 
reliable estimate of the uncertainty assuming that the priors
are reasonable and that the data are approximately Gaussian.
Resampling methods, such as the jackknife or the bootstrap,
can be used to check whether the distributions are Gaussian,
and they provide a simple check on statistical correlations
between different fit parameters.
Both procedures take many subsets of the data as
estimates of the original set; performing a fit on each subset
yields a distribution for each fit parameter from which an error
estimate can be made.  We employ the bootstrap method of 
resampling which requires
some modification in order to properly handle the contributions 
of the priors: as we show in Appendix~\ref{sec:fittingdetail}
one must randomly select new prior means $\mu_{\lambda_i}^b$
for each bootstrap fit \cite{Lepage:2001ym}.
Table \ref{tab:NRKS1link_mesps_820b1719m18} shows the results
of applying this bootstrap analysis to the heavy-light pseudoscalar
correlator computed with the 1-link staggered action on the
$8^3\times 20$ lattice.  These results can be compared
to those in Tables \ref{tab:efitvsk} and \ref{tab:afitvsk}.
We find both methods produce comparable error estimates.  For
ease of error propagation, we use bootstrap method to quote
uncertainties in the results presented below.

\section{Results}
\label{sec:results}

This section contains several results produced using
the methods proposed and described above.  The 
purpose of this study was to check the validity of this
proposal, so the results presented
below should not be construed as state-of-the-art calculations
to be used for phenomenology.  The results here show that
NRQCD-staggered calculations produce results comparable to
NRQCD-Wilson calculations -- central values agree and
statistical and fitting uncertainties are comparable --
but at a fraction of the computational cost.  A more complete
calculation of the $B$ spectrum and decay constant on finer,
unquenched configurations is
underway which will exploit the advantages of improved staggered
fermions to produce, we believe, the most accurate theoretical
computation of those quantities to date.

\subsection{Light hadron masses and dispersion relations}

As mentioned before we chose a value for the bare staggered mass
$m$ so that the ratio of the light pseudoscalar 
mass to the light vector mass would be somewhat near
the phenomenological value $\sqrt{2m_K^2 - m_\pi^2}/m_\phi = 0.673$.
We use the pseudo Goldstone boson correlator
$\tr |G_\chi(x;0)|^2$ to compute the pseudoscalar meson mass
and the correlator $(-1)^{x_k/a}\,\tr |G_\chi(x;0)|^2$ to compute
the vector meson mass.
These masses and their ratio are listed in Table~\ref{tab:pirho}
for the different lattices and actions.  Note that even
on the \ainv = 0.8 GeV lattice the
light hadron correlators from the AsqTad action 
do not suffer the contamination from the negative norm states
which affected the heavy-light correlators, as discussed in
Sec.~\ref{sec:fitting}.

\begin{figure}
\begin{center}
\epsfxsize=\hsize
\mbox{\epsfbox{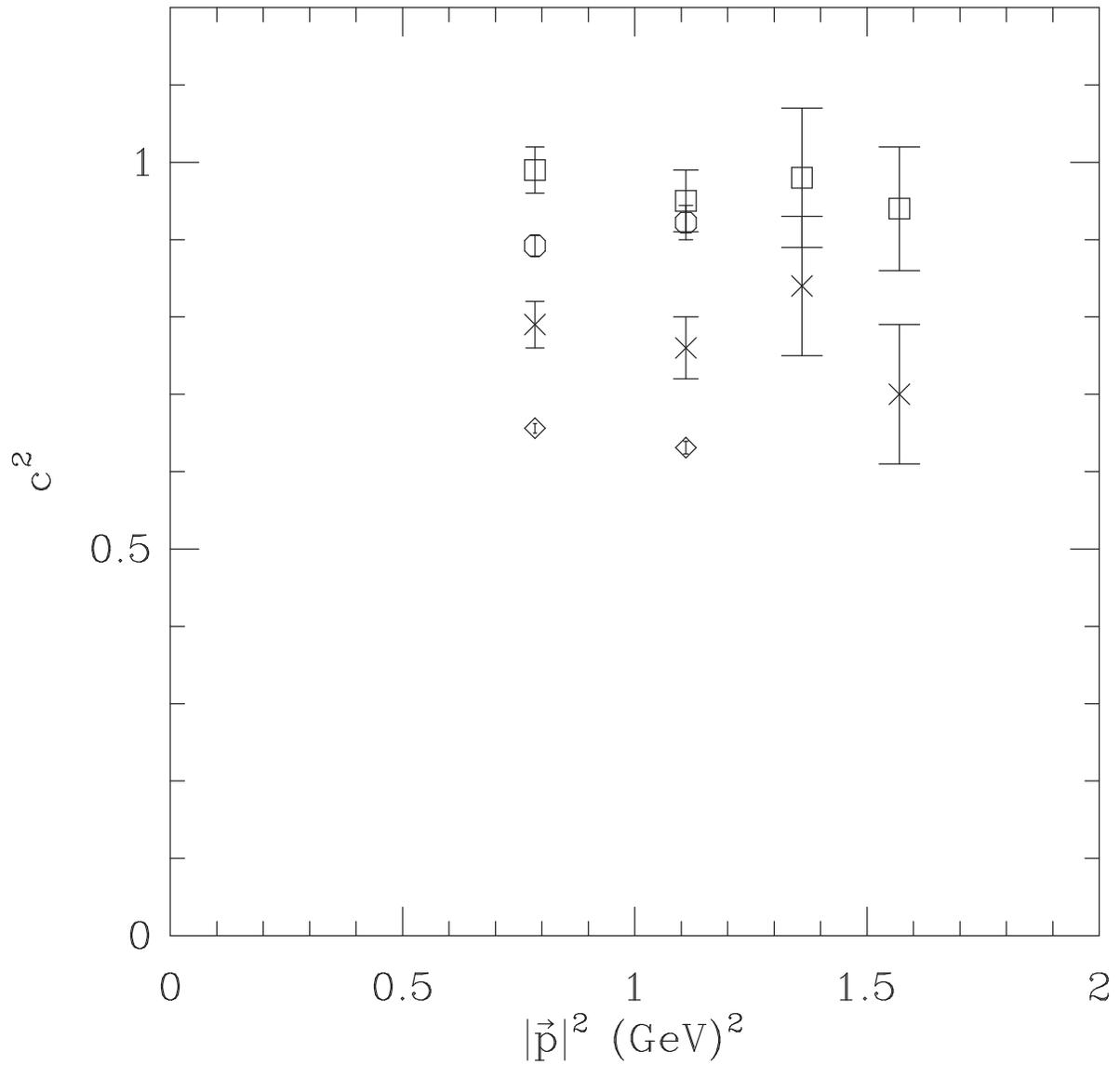}}
\caption{Pion speed-of-light squared vs.\ momenta 
on the coarse ($8^3\times 20$) lattice using several actions.
The 1-link (diamonds) and AsqTad (circles) results are ours,
compared to the clover (crosses) and D234 (squares) results
of \cite{Alford:1998yy}.}
\label{fig:lspeed}
\end{center}
\end{figure}

One measure of discretization effects is the dispersion
relation.  Specifically, we can compute the ``speed-of-light''
factor
\beq
c^2({\bf\bm{p}}) ~\equiv~ \frac{E^2(\VEC{p})-E^2(0)}{|\VEC{p}|^2}
\eeq
which should equal 1 in the absence of lattice artifacts.
The Naik term (\ref{eq:naik}) is responsible for subtracting the
$\order{a^2}$ uncertainties in $c^2({\bf\bm{p}})$ and its success
can be seen in the following results.
Table~\ref{tab:lspeed_iso}
lists the values of $c^2$ computed with several values of
momentum (averaged over all equivalent orientations in momentum
space).  On the coarser lattice ($\beta=1.719$) one can see that
using fat links does not improve $c^2$ much compared to the 1-link
action, but adding only the Naik term to the 1-link action results
in a significant improvement.  This is borne out on the finer 
lattice ($\beta=2.131$), where the AsqTad action has an improved $c^2$.
Fig.~\ref{fig:lspeed} shows comparison of $c^2(\VEC{p})$
between these results and
those for improved Wilson actions \cite{Alford:1998yy}.  
The AsqTad action has 
a better pion dispersion relation than the clover action, but not quite
as good as the D234 action.  

On the anisotropic lattices we use this quantity to tune the bare
parameter $c_0$ in (\ref{eq:anisoasqtad}); it is adjusted 
so that the pion speed-of-light
parameter $c^2(\VEC{p})\approx 1$.  Table~\ref{tab:lspeed_aniso}
lists the values of $c_0$ we used and the resulting computed values
of $c^2$ for several momenta.

\subsection{Finite momentum $B_s$ and its mass}

The energies, $E_{\rm sim}(\VEC{p})$,
extracted from correlation functions include an unknown but 
momentum independent shift due to the neglect of the
heavy quark rest mass, i.e.\
\beq
E_{\rm sim}(\VEC{p}) = E(\VEC{p}) - \Delta
\label{eq:esim}
\eeq
where $E(\VEC{p})$ is the physical energy.
In perturbation theory, the shift $\Delta$ is the difference
between the renormalized pole mass and the 
constant part of the heavy quark self-energy:
\beq
\Delta_{\rm pert} ~=~ Z_m M_0 - E_0 \, .
\eeq
Given $E_{\rm sim}(0)$ from a simulation,
the physical mass of a hadron can be computed
through
\beq
M_{\rm pert} ~\equiv~ E_{\rm sim}(0) + Z_M M_0 - E_0 
\label{eq:Mpert}
\eeq
where we attach the label ``pert'' to denote that 
the perturbative shift $\Delta_{\rm pert}$ was used.
For the finer isotropic lattice 
and the $1/M$ NRQCD action with $aM_0=5.0$ we find
\beq
Z_M M_0 - E_0 ~ = ~ M_0 - 0.890\,\alpha_s + M_0\,\order{\alpha_s^2}
\label{eq:ZMmE0}
\eeq 
The results obtained on the finer isotropic lattice, using the
AsqTad staggered action, give $M_{\rm pert}(B_s) = 5.55 \pm 0.45$ GeV
and $M_{\rm pert}(B_s^*) = 5.58 \pm 0.45$ GeV.
The numerical size of the $\order{\alpha_s^2}$ uncertainty 
is estimated by taking $\alpha_s\approx 0.3$, a typical value
for quenched lattices with these spacings, and assuming the coefficient
of the $\order{\alpha_s^2}$ term is 1 (times $M_0$ as indicated in
(\ref{eq:ZMmE0})).

The physical mass can also be calculated nonperturbatively,
using the dispersion relation
\beq
E^2(\VEC{p}) = M^2 + |\VEC{p}|^2 \,.  
\label{eq:heavydisprel}
\eeq
In order to cancel the unknown shift in (\ref{eq:esim}),
we consider $(E(\VEC{p}) - E(0))^2 = 
(E_{\rm sim}(\VEC{p}) - E_{\rm sim}(0))^2$, which we square
and solve for the mass
\beq
M_{\rm kin} ~\equiv~ 
\frac{|\VEC{p}|^2 - [E_{\rm sim}(\VEC{p})-E_{\rm sim}(0)]^2}
{2[E_{\rm sim}(\VEC{p})-E_{\rm sim}(0)]} \, .
\label{eq:mkin}
\eeq
When the mass is computed using (\ref{eq:mkin}), we call it
the kinetic mass, to distinguish it from the perturbative
result $M_{\rm pert}$.  Setting $|\VEC{p}| = 2\pi/12a = 0.52$ GeV, 
the kinetic masses on the finer isotropic lattice (with
the AsqTad light quark action) are $M_{\rm kin}(B_s) = 5.56 \pm 0.33$
GeV and $M_{\rm kin}(B_s^*) = 5.68 \pm 0.33$ GeV.

\begin{figure}
\begin{center}
\epsfxsize=\hsize
\mbox{\epsfbox{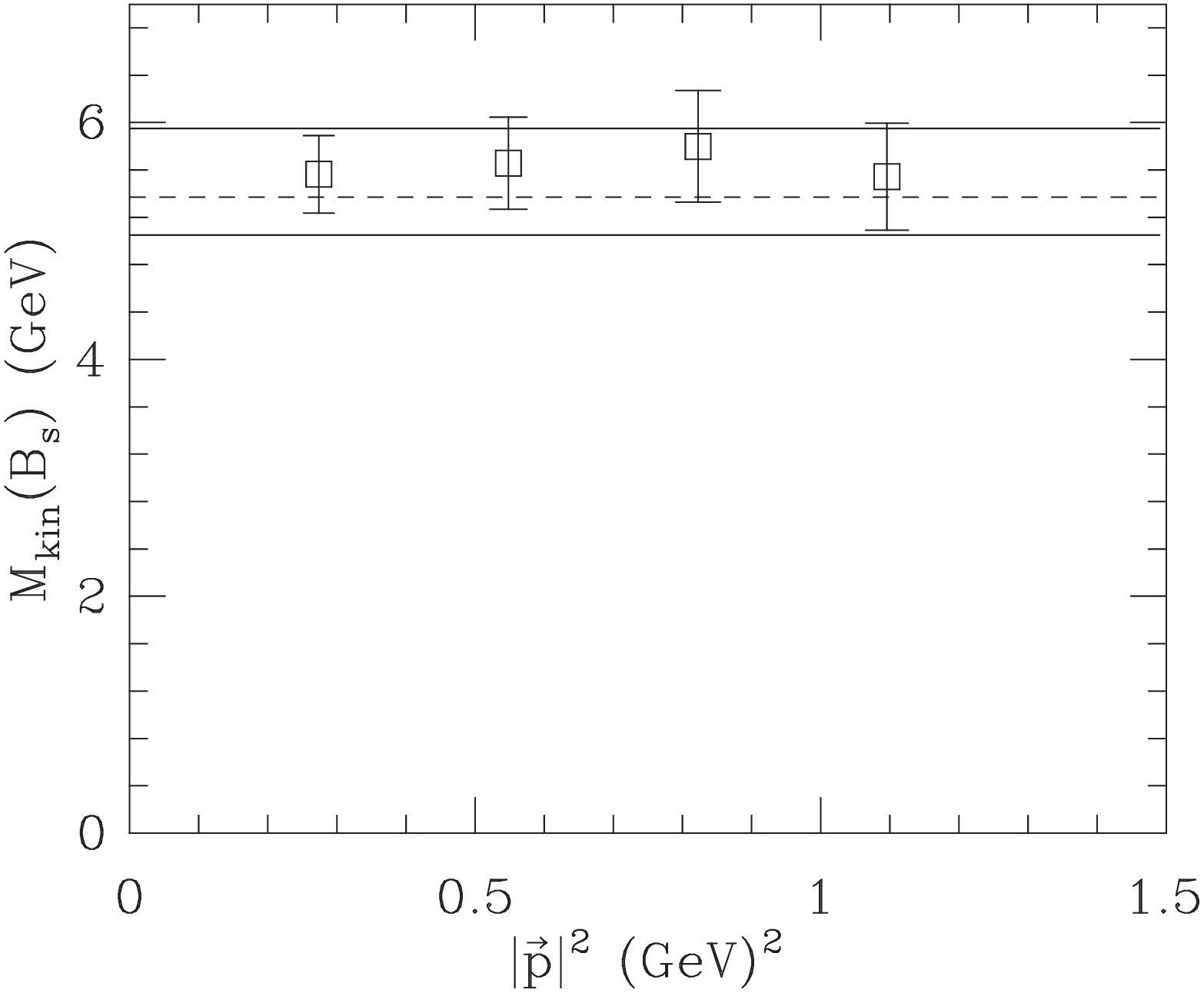}}
\caption{Kinetic mass for the $B_s$ meson on the $12^3\times32$ lattice
with the improved staggered action ($am=0.10$) and
the $1/M$ NRQCD action ($aM_0=5.0$).  Computed using (\ref{eq:mkin}).
The dashed line marks the experimental measurement $M_{B_s}= 5.37$ GeV,
and the solid lines show the range given perturbatively.}
\label{fig:mkinps_1232b2131m10M50}
\end{center}
\end{figure}

\begin{figure}
\begin{center}
\epsfxsize=\hsize
\mbox{\epsfbox{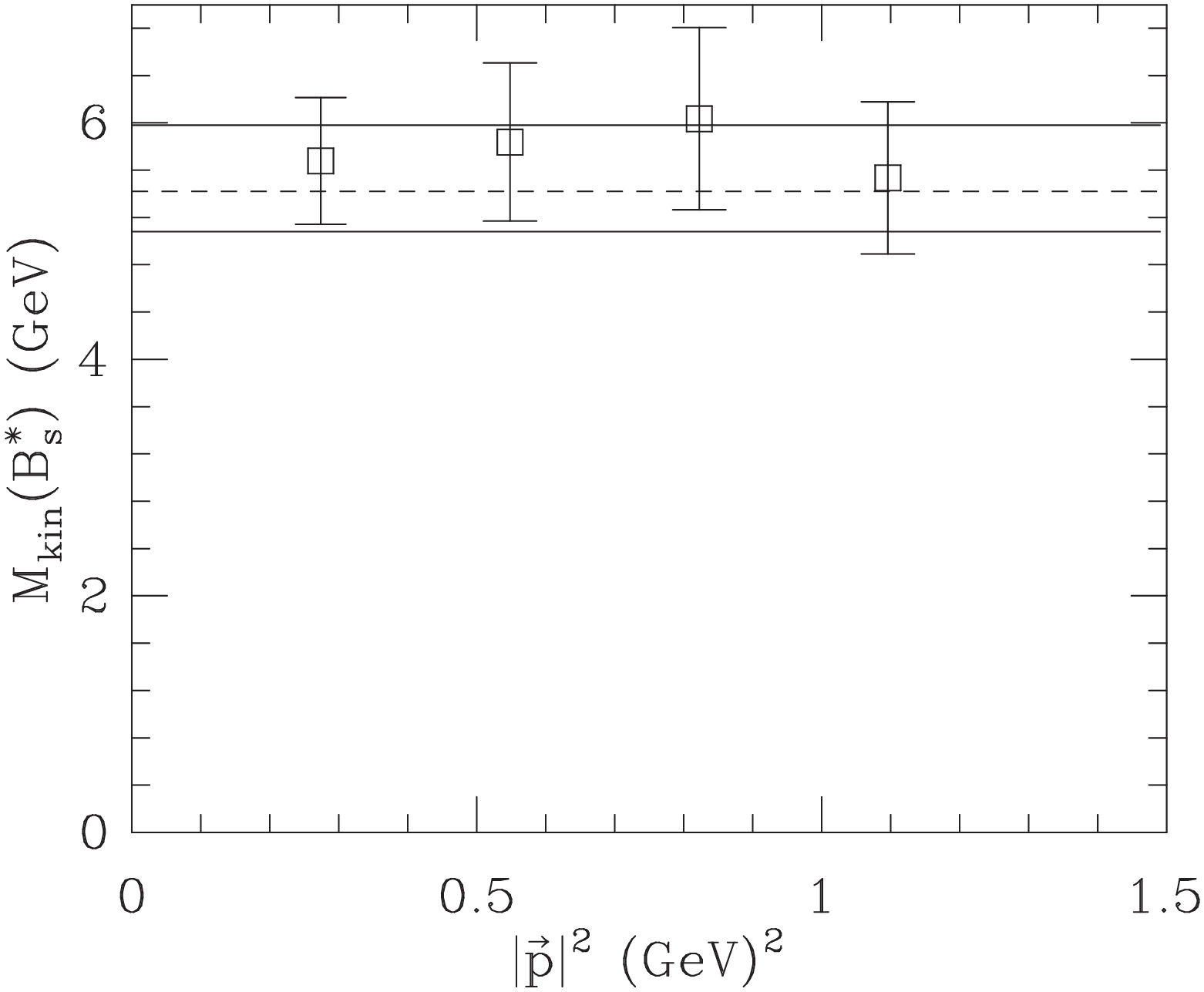}}
\caption{Kinetic mass for the $B_s^*$ meson on the $12^3\times32$ lattice
with the improved staggered action ($am=0.10$) and
the $1/M$ NRQCD action ($aM_0=5.0$).  Computed using (\ref{eq:mkin}).
The dashed line marks the experimental measurement $M_{B_s^*}= 5.42$ GeV,
and the solid lines show the range given perturbatively.}
\label{fig:mkinvk_1232b2131m10M50}
\end{center}
\end{figure}

Figs.~\ref{fig:mkinps_1232b2131m10M50} and 
\ref{fig:mkinvk_1232b2131m10M50} show the kinetic masses for
the $B_s$ and $B_s^*$ for several momenta.
We find excellent agreement between the perturbative and 
nonperturbative calculations of the mass.  Furthermore,
the consistency of the kinetic masses over several momenta
demonstrate that the combined NRQCD--improved staggered
formulation gives the correct dispersion relation for
$B_s$ and $B_s^*$ up to $|\VEC{p}| = \pi/3a = 1.1$ GeV.
One should not
put too much weight on any agreement or disagreement between
the calculation and experiment, given that the calculation
is quenched, the lattice spacing not precisely determined, and
the quark masses not precisely tuned.  


\subsection{Mass splittings in the $B_s$ system}

Since the shift $\Delta$ between simulation energy and 
the physical energy (Eq.\ \ref{eq:esim})
is entirely due to the NRQCD action,
it is universal for all bound states with the heavy quark.
Therefore, we can compute mass splittings much more precisely
than suggested by the uncertainties in $M_{\rm kin}$ and 
$M_{\rm pert}$.  The splittings we compute on various
lattices  which correspond
to the $B_s$ system are given in Table~\ref{tab:Bs_splittings};
below are a few remarks concerning the different calculations.

\begin{figure}
\begin{center}
\epsfxsize=\hsize
\mbox{\epsfbox{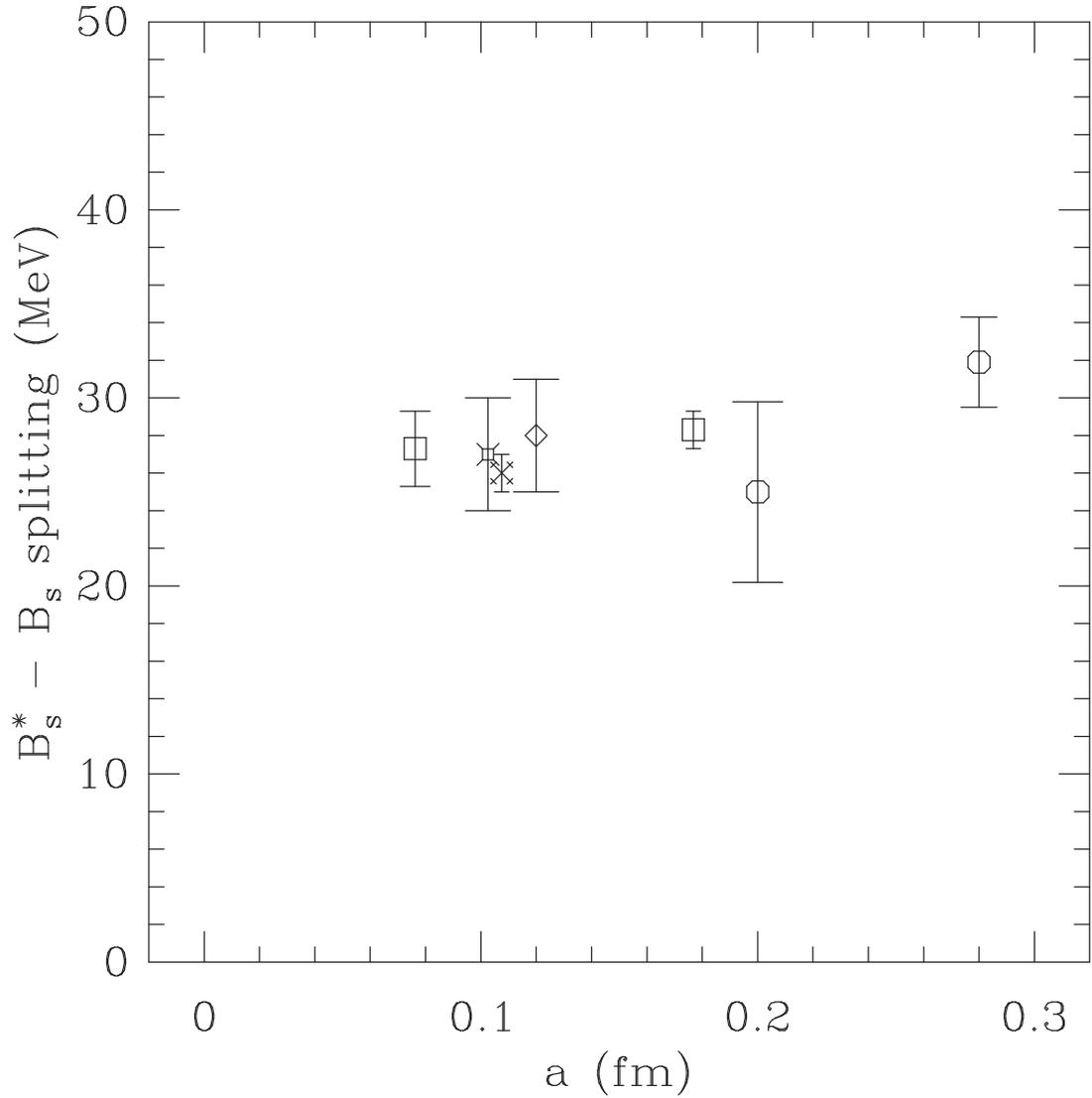}}
\caption{The hyperfine splitting between $B_s$ and $B_s^*$ computed
on quenched lattices.  
Our NRQCD-staggered results (from the isotropic lattices) are the circles.
The squares come from \cite{Hein:2000qu}, the fancy
square from \cite{AliKhan:1999yb}, and the
diamond from \cite{Ishikawa:1999xu} which all used 
an NRQCD-clover action, and the fancy cross comes from
an NRQCD-D234 calculation \cite{Lewis:2000sv}.
All tune the heavy quark mass
as in this work (see text for elaboration).
For comparison, the experimental measurement is 
$47.0\pm 2.6$ MeV \cite{Hagiwara:2002pw}.
Error bars are statistical only.  }
\label{fig:BstarBsplit}
\end{center}
\end{figure}

The hyperfine splitting $M_{B_s^*} - M_{B_s}$ is the most 
straightforward to compute since it is the difference between
the $E_{\rm sim}$ for the non-oscillating ground states of
the vector and pseudoscalar correlators.  The results are
comparable to previous quenched lattice studies;
Figure~\ref{fig:BstarBsplit} shows our quenched results 
on the 2 isotropic lattices compared to results published in 
Refs.~\cite{AliKhan:1999yb,Ishikawa:1999xu,Hein:2000qu,Lewis:2000sv}.  
This splitting was also computed using NRQCD in 
Ref.~\cite{Lewis:1998ka}, but they have
different systematic errors caused by the quenched approximation, 
specifically they set the bare bottom quark mass by tuning
the $\Upsilon$ mass, instead of a heavy-light mass.
Our error bars are larger than those for most other results for
two reasons.  The first is simply that this work is based on 200
configurations compared to 300 \cite{Ishikawa:1999xu}, 
278 and 212 \cite{Hein:2000qu}, and 2000 \cite{Lewis:2000sv}
(Ref.~\cite{AliKhan:1999yb} used 102 configurations).
The second is that the Bayesian curve fitting method includes as part
of the quoted uncertainty an estimation of the error due to 
excited state contamination, in contrast to the single exponential fits
used in previous work.

Quenched results have an inherent ambiguity 
depending on which physical quantities are used to
set the lattice spacing and bare quark masses.  
Preliminary results  on unquenched lattices
indicate that the inclusion of sea quarks
yields a unique scale and bottom quark masses \cite{Gray:2002vk}
and give a $B_s^* - B_s$ splitting \cite{Wingate:2002ry} 
consistent with the experimental measurement
$M_{B_s^*} - M_{B_s} = 47.0 \pm 2.6$ MeV \cite{Hagiwara:2002pw}.

The $L=1$, or ``P-wave'', states $B_{s0}^*$ and $B_{s1}$ have the 
same quantum numbers as the oscillating states in the pseudoscalar and vector
correlators, respectively.  The fact that $E_{\rm sim}$ for these
states can be computed using the same correlator data as
the $L=0$ states should be another advantage over formulations
with Wilson-like quarks.  In practice, however, it appears that
the coupling of these states to the local-local correlator is rather small
and consequently the fitting uncertainties for these splittings
are large.  Smeared sources and sinks for both heavy
and light quark propagators should be explored as methods for
amplifying the coupling to the $P$-wave states.
In Table~\ref{tab:Bs_splittings} we list some
combinations of splittings.  

\subsection{Decay constant}

The heavy-light decay constants are defined through the matrix
element of the electroweak axial vector current
\beq
\langle 0 | A_0 | B_s \rangle ~=~ 
\langle 0 | \; \overline{q}\, \gamma_5\gamma_0 \,b\; | B_s \rangle
~=~ f_{B_s} M_{B_s} \, .
\label{eq:fBdef}
\eeq
The fields in the current above are those defined in the 
Standard Model, so a matching must be performed between
them and the fields of our lattice action.
The continuum heavy quark field $b$ is related to the nonrelativistic
field $\phi$ through the Foldy-Wouthuysen-Tani transformation
\beq
b ~=~ \left( 1 - \frac{\bm{\gamma\cdot\nabla}}{2 M_0}
~+~ \order{\left(\frac{\Lambda_{\rm QCD}}{M}\right)^2} \right) Q
\eeq
where 
\beq
Q ~=~ \left( \begin{array}{c} \phi \\ 0 \end{array}\right) \,.
\eeq
Expanding the QCD axial-vector current in terms of 
NRQCD operators up to $\order{\Lambda_{\rm QCD}/M}$
and at $\order{\alpha_s}$ in perturbation theory yields a
combination of the three operators
\bea
J_0^{(0)} & = & \overline{\Psi} \, \gamma_5\gamma_0 \, Q 
\label{eq:j00} \\
J_0^{(1)} & = & -\frac{1}{2M_0}\overline{\Psi}\, \gamma_5\gamma_0 \,
{\bf\bm{\gamma\cdot\nabla}} \,Q \\
J_0^{(2)} & = & \frac{1}{2M_0}\overline{\Psi} \,
\bm{\gamma\cdot\stackrel{\leftarrow}{\nabla}}\, \gamma_5\gamma_0 \,Q \, .
\eea
The operator equation is then written as 
\beq
A_0 ~\doteq~ (1+\alpha_s \rho_0) J_0^{(0)} 
+ (1+\alpha_s \rho_1) J_0^{(1)}
+ \alpha_s \rho_2 J_0^{(2)} \, .
\label{eq:j0j1j2}
\eeq
The symbol $\doteq$ is meant to imply that matrix elements of
the operators on the left and right hand sides are equal,
up to whatever order in the effective theory we are working.
Since we are neglecting terms of order $\alpha_s\Lambda_{\rm QCD}/M$
the terms proportional to $\rho_1$ and $\rho_2$ are dropped from
our analysis.  The relation we use to do the matching is
\beq
 A_0 ~\doteq~ (1+\alpha_s\wtilde{\rho}_0) J_0^{(0)}  \;+\;  
J_0^{(1,{\rm sub})} 
\label{eq:axialvec}
\eeq
where the $1/(aM)$ power law mixing of $J_0^{(1)}$ with $J_0^{(0)}$
is absorbed at one-loop level into a subtracted $\Lambda_{\rm QCD}/M$ 
current \cite{Collins:2000ix}
\beq
J_0^{(1,{\rm sub})} ~\equiv~ J_0^{(1)} - \alpha_s \zeta_{10} J_0^{(0)} \, ,
\label{eq:j01sub}
\eeq
 and $\wtilde{\rho}_0 - \zeta_{10}=\rho_0$. 

Since the heavy spinor obeys $\gamma_5\gamma_0 Q = \gamma_5 Q$,
the matrix element ${\langle 0 | J_0^{(0)} | B_s\rangle}$
is related simply to the ground state amplitude of the pseudoscalar
heavy-light correlator $C_{B}^{(2)}(\VEC{p}=0,t)$.  Let us denote 
this amplitude by $C_{00}$, then 
\beq
C_{00} ~=~ \frac{|\langle 0 | J_0^{(0)} | B_s\rangle|^2}{2 M_{B_s}} \, .
\eeq
To get the $\Lambda_{\rm QCD}/M$ current matrix element
we compute correlators where we put $J_0^{(1)}$
at the sink. Let us denote the ground state amplitude of this
correlator by $C_{10}$, then
\beq
C_{10} ~=~ \frac{\langle 0 | J_0^{(1)} | B_s\rangle
\langle B_s | J_0^{(0),\dagger} | 0 \rangle}{2 M_{B_s}} \, .
\eeq

\begin{figure}
\begin{center}
\epsfxsize=\hsize
\mbox{\epsfbox{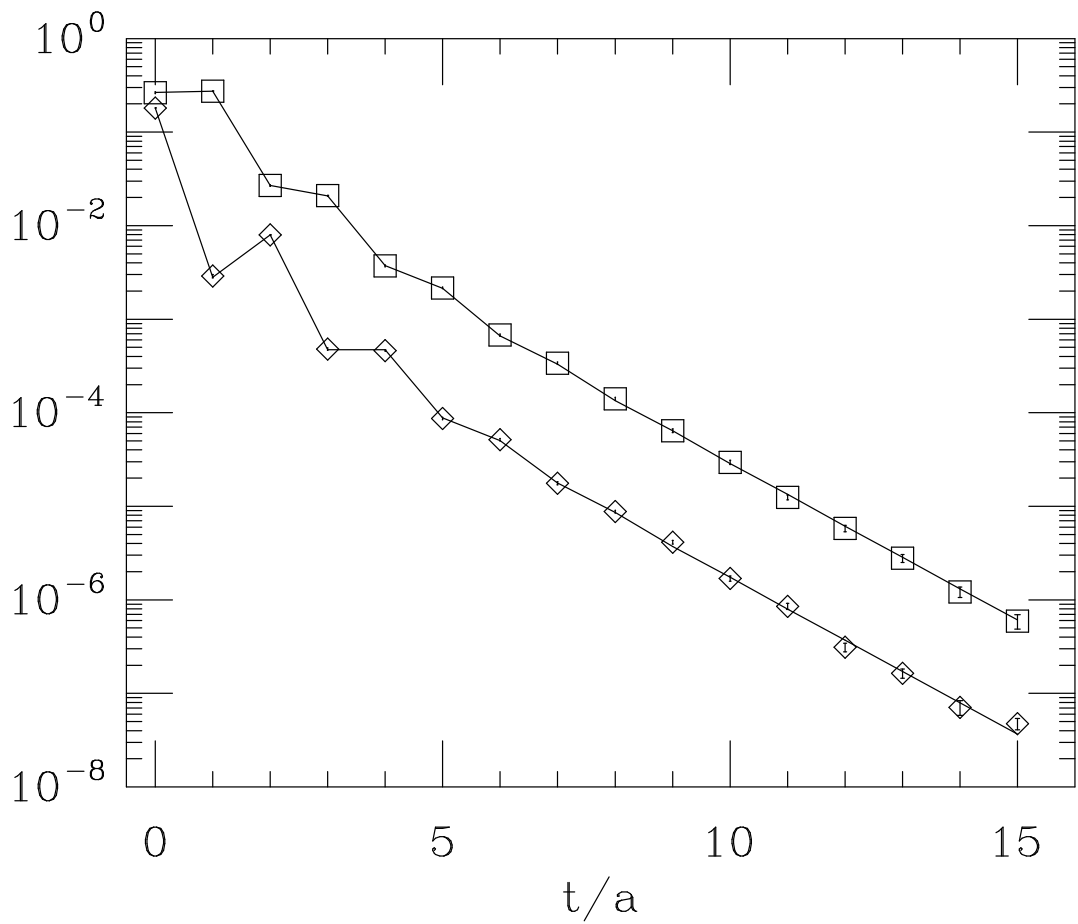}}
\caption{Correlators $\langle J_0^{(0)}(t)J_0^{(0),\dagger}(0)\rangle$
(squares) and
$-\langle J_0^{(1)}(t)J_0^{(0),\dagger}(0)\rangle$ (diamonds)
necessary for calculation of $f_{B_s}$ through $\Lambda_{\rm QCD}/M$.
Computed on the isotropic $12^3\times 32$, $1/a = 1.0$ GeV lattice.}
\label{fig:j0ratio}
\end{center}
\end{figure}

As mentioned before, we concentrate on the quenched $12^3\times 32$
lattice which is closest to the target unquenched configurations,
albeit coarser.  
Fits to these correlators, shown in Fig.~\ref{fig:j0ratio},
yield the bootstrapped ratio
\beq
\frac{\langle 0 | J_0^{(1)} | B_s\rangle}
{\langle 0 | J_0^{(0)} | B_s\rangle} ~=~ -0.064 \pm 0.005 ~{\rm (stat).}
\label{eq:unsubJ01ratio}
\eeq
The $1/M$ NRQCD action is used for this calculation, for which
we compute (with $aM_0=5.0$) $\wtilde{\rho}_0 = 0.208 \pm 0.003$ and
$\zeta_{10} = -0.0997$.  Performing the subtraction (\ref{eq:j01sub})
we find
\beq
\frac{\langle 0 | J_0^{(1,{\rm sub})} | B_s\rangle}
{\langle 0 | J_0^{(0)} | B_s\rangle} ~=~ -0.034 \pm 0.004 ~{\rm (stat).}
\label{eq:subJ01ratio}
\eeq
This ratio can be compared to other lattice formulations;
it is the ``physical'' $\Lambda_{\rm QCD}/M$ correction
to $J_0^{(0)}$ with $1/a$ power law effect subtracted at
the one-loop level.  The 3.4(4)\% correction we find on the
\ainv~$= 1.0$ GeV lattice is
in excellent agreement with the 3-5\% corrections found
using the NRQCD and clover actions on lattices with
inverse spacings from 1.1 -- 2.6 GeV \cite{Collins:2000ix}.
Note that even on the finest lattice in Ref.~\cite{Collins:2000ix}, 
where power law contributions are the largest, the one-loop subtraction
takes ${\langle 0 | J_0^{(1)} | B_s\rangle}/
{\langle 0 | J_0^{(0)} | B_s\rangle} = -14\%$ to 
${\langle 0 | J_0^{(1,{\rm sub})} | B_s\rangle}/
{\langle 0 | J_0^{(0)} | B_s\rangle} = -4\%$, in agreement
with calculations on coarser lattices.  Given the present
agreement between our result and that of Ref.~\cite{Collins:2000ix},
we can expect a similarly successful subtraction in our
ongoing calculation with the unquenched MILC ensemble.

Applying (\ref{eq:axialvec}) and (\ref{eq:fBdef})
gives the quenched result
\beq
f_{B_s} ~=~ 225 \pm 9 {\rm (stat)} \pm 20 {\rm (p.t.)} 
\pm 27 {\rm (disc.)}~{\rm MeV.} 
\eeq
The 20 MeV perturbative uncertainty is the 
estimate of the $\order{\alpha_s^2}$
error in (\ref{eq:axialvec}), obtained by taking
$\alpha_s\approx 0.3$ and a coefficient equal to 1.
The other perturbative uncertainties, due to one-loop corrections
to the coefficients in the action and in the operator matching,
are $\order{\alpha_s\Lambda_\mathrm{QCD}/M}$ which is estimated to
be 2.4\%,
assuming $\Lambda_\mathrm{QCD}$ = 400 MeV (and using $M_0 = 5.0$ GeV).
The 27 MeV discretization error is our estimate of the
$\order{\alpha_s a \Lambda_{\rm QCD}}$ error in the 
current $J_0^{(0)}$ (\ref{eq:j00});
again we assume the leading correction term comes
with a coefficient of order 1.
This error may be reduced to 
$\order{\alpha_s (a \Lambda_{\rm QCD})^2}$ by improving $J_0^{(0)}$
{\it\`a la} Symanzik,
which requires calculation of $\rho_1$ and $\rho_2$ in
(\ref{eq:j0j1j2}) \cite{Shigemitsu:1998xf,Morningstar:1998ep}.
Finally, note that we have included the $\order{\alpha_s/aM}$ power-law
correction; we would have estimated this to be a 6\% effect, but
it was calculated to be 3\% (compare \ref{eq:unsubJ01ratio} and
\ref{eq:subJ01ratio}). 
Given those uncertainties, we find agreement
with the recent quenched world average 
$f_{B_s} ~=~ 200\pm 20$ MeV \cite{Ryan:2001ej}.


\section{Conclusions}
\label{sec:concl}

We believe the methods outlined within this paper provide the
quickest route to accurate calculations of $B$ meson masses
and decay constants on realistic unquenched lattices.  Improved
staggered fermions have several advantages over Wilson-like
fermions and are far less expensive to simulate than domain wall
or overlap fermions.  The equivalence between naive and staggered
fermions greatly simplifies the construction of operators
which couple to states of interest.  The fact that the
NRQCD action does not have a doubling symmetry leads to taste-changing
suppression in heavy-light mesons, avoiding the ambiguities
of the light staggered hadrons.

We have presented results on several types of lattices, the most
important being the finer of the two isotropic lattices since it
is most similar to the unquenched MILC lattices.  The results from
these simulations have no unpleasant surprises: they agree with
results produced by previous quenched simulations.  Therefore, we
can trust this formulation when it is used in parts of parameter
space inaccessible to other formulations.

\begin{acknowledgments}
We are grateful to Kerryann Foley for computing the static quark
potential on the $12^3\times 32$ lattice and to Quentin Mason
for providing Feynman rules for the AsqTad action.
Simulations were performed at the 
Ohio Supercomputer Center and at NERSC; some code was derived from the
public MILC Collaboration code 
(see http://physics.utah.edu/$\sim$detar/milc.html).
This work was supported in part by the U.S.\ DOE, NSF, PPARC, and NSERC.
J.S.\ and M.W.\ appreciate the hospitality of the Center for Computational
Physics in Tsukuba where part of this work was done.
\end{acknowledgments}

\appendix

\section{Formalism details}
\label{sec:formaldetail}

In this Appendix we present a more detailed analysis of the heavy-light
operators used in the numerical calculation.  

Since naive fermions have 16 taste degrees-of-freedom, there is 
the possibility of forming 16 different $B$ mesons labeled by the 
light taste index $g$, i.e. $B_g$.  The general choice for 
a $B_g$ meson interpolating 
heavy-light operator takes on the form
\beq
\label{bmg}
{\cal W}_{B_g}(x) = \hpsid(x) \gamma_5 M_g e^{i\pi_g \cdot x}
 \Psi(x)
\eeq
The 16 different operators lead to degenerate states, since they are related 
by the symmetry transformation (\ref{dsymm}). It is sufficient to work 
with just one of the 16 choices to extract all the relevant physics. 
 In our simulations 
we usually use the simplest choice $g = \emptyset$, i.e.\ Eq.~(\ref{bm}). 
Any other choice would have served equally well. For instance, consider the 
case $g = \mu_j$ with $\mu_j$ equal to one of the spatial directions and 
carry out a sum over spatial sites.
Eq.~(\ref{bmg}) then becomes
\beq
\label{bmgj}
a^3 \, \sum_{\xv}{\cal W}_{B_j}(x) = 
a^3 \, \sum_{\xv}\hpsid(x) \gamma_5 M_j e^{i\pi_j \cdot x} \Psi(x)
 = a^3 \, \sum_{\xv}\hpsid(x) i \gamma_j e^{i\pi_j \cdot x} \Psi(x).
\eeq
One sees that 
the zero spatial momentum $B_j$ meson operator is identical to 
an operator one would superficially (and incorrectly) 
associate with a $B^*$ meson of 
polarization ``$j$'' with momentum $\pi/a$ in the $j^{th}$ direction. 
The correct interpretation of (\ref{bmgj}) is that it represents a 
zero momentum pseudoscalar heavy-light meson. This will become more evident 
when we look at the operator 
${\cal W}_{B_g}(x)$ in momentum space.  
 We have verified that the RHS of 
(\ref{bmgj}) gives identical correlators, configuration-by-configuration, 
to (\ref{bm}) (the latter summed over space).  (In fact, the symmetries 
of (\ref{dsymm}) provide excellent tests of one's simulation codes.)
Therefore, it is sufficient to work with just one type of $B$ meson operator, 
e.g.\ with just (\ref{bm}).  

In order to delve further into the Lorentz quantum number and taste content 
of the interpolating operators ${\cal W}_{B_g}(x)$ we will 
look at this operator in momentum space.  
In terms of the ``tilde'' fields (\ref{eq:3dtilde}) one has
(we take the 
case where $g$ does not include a temporal component;
the latter case can be discussed in a completely analogous way)
\bea
\label{wmb}
a^3 \, \sum_{\xv} \wcal_{B_g}(\xv,t) & = &
\int_{\kv,D_s} \overline{\tilde{\psi}_H}(\kv,t)
 \, \gamma_5 M_g \, \tilde{\psi}(\kv+\piv_g,t)  \nl
& = &\sum_{g_s \in G_s} \int_{\kv,D_{s,\emptyset}} \overline{\tilde{\psi}_H}
(\kv+\piv_{g_s},t)
 \, \gamma_5  M_g \, \tilde{\psi}(\kv + \piv_g + \piv_{g_s},t)
\eea
We extract the taste content of this bilinear by writing 
\begin{eqnarray}
\tilde{\psi}(\kv + \piv_g + \piv_{g_s},t) &=& \ktint e^{ik_0t} 
\psi(\kv + \piv_g + \piv_{g_s},k_0) \nl
&=& \hktint e^{ik_0t} \left[ \psi(\kv+\piv_g+\piv_{g_s},k_0) + 
(-1)^t \psi(\kv + \piv_g+\piv_{g_s},k_0 + \pi_{g_t}) \right]  \nl
&=& \hktint e^{ik_0t} \left[ M_{g_sg}^\dagger q^{g_sg}(\kv,k_0) + 
(-1)^t M^\dagger_{g_tg_sg} q^{g_tg_sg}(\kv,k_0) \right]  
\end{eqnarray}
so that
\begin{eqnarray}
\label{bg}
a^3 \, \sum_{\xv} \wcal_{B_g}(\xv,t) & = &
 \sum_{g_s \in G_s} \int_{\kv,D_{s,\emptyset}} 
  \hktint e^{ik_0t} \nl
 & & 
 \left\{\overline{\tilde{\psi}}_H(\kv + \piv_{g_s},t) \,  \gamma_5
M_{g} 
 \left[ M_{g_sg}^\dagger q^{g_sg}(\kv,k_0) + 
(-1)^t M^\dagger_{g_tg_sg} q^{g_tg_sg}(\kv,k_0) \right]   \right\}
\end{eqnarray}
Since there is no doubling symmetry for the heavy quark action, the field  
$\overline{\tilde{\psi}}_H(\kv + \piv_{g_s},t) $, for $\piv_{g_s} \ne 
\piv_{\emptyset}$, represents a heavy quark with large spatial momentum.
Consequently, even though the operator in (\ref{bg}) couples to 
zero momentum meson states, the states corresponding to
$g\ne\emptyset$ are very energetic.  This is precisely the important
difference between studying heavy-light and light-light mesons
with light staggered quarks.  

We will estimate the effect of the $g_s \neq \emptyset$ sectors 
below, however the lowest energy state, and consequently the dominant
 contributions to a  $ \wcal_{B_g}(\xv,t)$
 correlator, will come from the region 
$g_s = \emptyset$ in the sum $\sum_{g_s}$.
\begin{eqnarray}
\label{bg0}
a^3 \, \sum_{\xv} \wcal_{B_g}(\xv,t) & \rightarrow &
 \int_{\kv,D_{s,\emptyset}} 
  \hktint e^{ik_0t} \nl
 & & 
 \left\{\overline{\tilde{\psi}}_H(\kv,t) \,  \gamma_5 \,
 \left[ q^{g}(\kv,k_0) + 
(-1)^t M^\dagger_{g_t} q^{g_tg}(\kv,k_0) \right]   \right\}
\end{eqnarray}
The non-oscillating contribution is the $B_g$ meson of taste
$g$.  Its parity partner is a $0^+$ meson, 
usually called the $J=0$ $P-$state.  It is remarkable that both 
$S-$ and $P-$states can be obtained from a single correlator.
Note also that the combination 
$ \overline{\tilde{\psi}}_H(\kv,t) \,  \gamma_5 \,
q^{g}(\kv,k_0) $, with its obviously pseudoscalar Lorentz structure 
holds for all tastes $g$, i.e. for trivial and 
nontrivial $M_g$ in (\ref{bmg}).

We discuss next those terms omitted upon going from 
(\ref{bg}) to (\ref{bg0}).  Take for instance the contribution from 
$g_s \rightarrow gl \equiv \mu_l$ with ``$l$'' equal to one of the
spatial directions.  The non-oscillatory term becomes
\beq 
\label{bmgl}
\overline{\tilde{\psi}}_H(\kv + \piv_{gl},t) \,  \gamma_5
M_{gl} \, q^{gl \, g}(\kv,k_0) = 
\overline{\tilde{\psi}}_H(\kv + \piv_{gl},t) \,  i \gamma_l
 \, q^{gl \, g}(\kv,k_0).
\eeq
One sees that the Lorentz structure is that of a $1^-$ particle.  
However, since the heavy quark has very high momentum and no doublers,
this intermediate state is highly virtual.  Such states would appear
in fits to correlation functions as extra structure at energies
of order $\Delta E \approx 1/(M a^2)$.  
These lattice artifacts can also affect low energy states through 
loops; their effects can be estimated perturbatively and are part of the 
$\order{\alpha_s a^2}$ errors inherent in the action.
Such errors can be removed, if need be, by perturbatively improving
the action further, but there is little evidence that they are
important at practical values of the lattice spacing.

\section{Discrete derivatives and field strengths}
\label{sec:derivatives}

Here we write explicitly the higher order tadpole-improved derivatives 
and improved field strength tensor used in the fermion actions.
\begin{eqnarray}
\nabla^{(2)}_\mu\, \Psi(x) &=&  \frac{1}{u_\mu} \,
 [ U_\mu(x) \Psi(x+a_{\mu})
\; + \; U^\dagger_\mu(x - a_{\mu}) \Psi(x - a_{\mu}) ] \nl
&& \; - \; 2\,\Psi(x) \\
\nabla^{(3)}_\mu\, \Psi(x) &=& 
\frac{1}{2} \, \frac{1}{u^2_\mu} \, [ U_\mu(x) U_\mu(x +  a_{\mu})
 \Psi(x+ 2a_{\mu}) \nl
&&\; - \; U^\dagger_\mu(x - a_{\mu}) U^\dagger_\mu(x - 2 a_{\mu})
 \Psi(x - 2a_{\mu}) ]  \nl
 & & - \, \frac{1}{u_\mu} \, [ U_\mu(x) \Psi(x+a_{\mu})
\; - \; U^\dagger_\mu(x - a_{\mu}) \, \Psi(x - a_{\mu}) ]  \\
\nabla^{(4)}_\mu\, \Psi(x) &=& 
 \frac{1}{u^2_\mu} \, [ U_\mu(x) U_\mu(x +  a_{\mu})
 \Psi(x+ 2a_{\mu}) \nl
&&\; + \; U^\dagger_\mu(x - a_{\mu}) U^\dagger_\mu(x - 2 a_{\mu})
 \Psi(x - 2a_{\mu}) ]  \nl
 & & -\,4 \, \frac{1}{u_\mu} \, [ U_\mu(x) \Psi(x+a_{\mu})
\; + \; U^\dagger_\mu(x - a_{\mu}) \, \Psi(x - a_{\mu}) ] \nl
&& \; + \; 6\,\Psi(x) 
\end{eqnarray}
The covariant derivatives acting on link matrices are defined as follows:
\begin{eqnarray}
\frac{1}{u_\nu}\nabla_\mu \,U_\nu(x) &=& \frac{1}{u_\mu^2 u_\nu} 
\,\Big[ U_\mu(x) \, U_\nu(x+a_\mu) \, U^\dagger_\mu(x+a_\nu) \nl 
&& \;-\;
U^\dagger_\mu(x-a_\mu) \, U_\nu(x-a_\mu) \, U_\mu(x-a_\mu+a_\nu) \Big]
\\
\frac{1}{u_\nu}\nabla^{(2)}_\mu\, U_\nu(x) &=& \frac{1}{u_\mu^2 u_\nu} 
\,\Big[ U_\mu(x) \, U_\nu(x+a_\mu) \, U^\dagger_\mu(x+a_\nu) \nl 
&& \;+\;
U^\dagger_\mu(x-a_\mu) \, U_\nu(x-a_\mu) \, U_\mu(x-a_\mu+a_\nu) \Big]
\nl && - \frac{2}{u_\nu}\, U_\nu(x)
\end{eqnarray}
The field strength operator $F_{\mu\nu}(x)$
is constructed from the so-called
clover operator $\Omega_{\mu\nu}(x)$
\begin{eqnarray}
  F_{\mu\nu}(x) &=& {1\over 2i}
\left( \Omega_{\mu\nu}(x)-\Omega^\dagger_{\mu\nu}(x)\right), \nonumber\\
\Omega_{\mu\nu}(x) &=&\!{1\over 4u_\mu^2 u_\nu^2}\!\sum_{{\lbrace(\alpha,\beta)
   \rbrace}_{\mu\nu}}\!\!\!\!U_\alpha(x)U_\beta(x\!+\!a_\alpha)
   U_{-\alpha}(x\!+\!a_\alpha\!+\!a_\beta)U_{-\beta}(x\!+\!a_\beta)
\label{fieldstrength}
\end{eqnarray}
where the sum is over 
$\lbrace(\alpha,\beta)\rbrace_{\mu\nu} = \lbrace (\mu,\nu),
(\nu,-\mu), (-\mu,-\nu),(-\nu,\mu)\rbrace$ for $\mu\neq\nu$ and 
$U_{-\mu}(x+a_{\mu}) \equiv U^\dagger_\mu(x)$.  
The $\order{a^2}$ improved field strength tensor is
\begin{eqnarray}
\tilde{F}_{\mu\nu}(x) ~ = ~ \frac{5}{3}F_{\mu\nu}(x) 
 & - & \frac{1}{6} \Bigg[\,\frac{1}{u_\mu^2} 
(\,U_\mu(x)F_{\mu\nu}(x+a_\mu)U^\dagger_\mu(x)  \nl
& + & U^\dagger_\mu(x-a_\mu)
F_{\mu\nu}(x-a_\mu)U_\mu(x-a_\mu)\,) 
 -  (\mu \leftrightarrow \nu)\,\Bigg] \nl
& + & \frac{1}{6} \, ( \frac{1}{u_\mu^2} + \frac{1}{u_\nu^2} 
- 2 ) \, F_{\mu\nu}(x).
\end{eqnarray}

\section{Fitting details}
\label{sec:fittingdetail}

In this Appendix we give a pedagogical discussion of the
constrained curve fitting proposed in \cite{Lepage:2001ym}.

Recall that the standard fitting procedure is to 
minimize the $\chi^2$ function, or equivalently, to maximize 
the likelihood of the data, $C(t)$, given a set of fit parameters.
The likelihood probability is given, up to a normalization constant, by
\beq
\prob( C(t) | f(t;{\bm \lambda}), I) ~\propto~
\exp\left(-\frac{\chi^2}{2}\right) 
\label{eq:likelihood}
\eeq
where $I$ represents any unstated assumptions.  Explicitly,
\beq
\chi^2 ~=~ \sum_{t,t'}\Big(\vev{C(t)} - f(t;{\bm \lambda})\Big)\, 
K_{t,t'}^{-1} \, \Big(\vev{C(t')} - f(t';{\bm \lambda})\Big) \, .
\label{eq:chisq}
\eeq
The correlation matrix, $K$, is constructed to take into
account correlations between $C(t)$ and $C(t')$:
\beq
K_{t,t'} ~\equiv~ \frac{1}{N-1} \; \left\langle
\Big(C(t) - \vev{C(t)}\Big)\Big(C(t') - \vev{C(t')}\Big)\right\rangle\, .
\label{eq:corrmat}
\eeq
with $N$ equal to the number of measurements.

Usually one cannot include enough terms in the fit to account
for excited state contributions before the algorithm for
minimizing $\chi^2$ breaks down.  
The minimization algorithm diverges as it searches in 
directions of parameter space which are unconstrained by the data.
In the past the solution has been to limit the number of fit terms,
then discard data by including $C(t)$ for $t\ge t_{\rm min} > 0$;
the optimal value of $t_{\rm min}$ is selected by a combination
of looking for $\chi^2/{\rm DoF} = 1$, maximizing the confidence level
(Q factor), and observing plateaux in effective masses.  A major
weakness of this procedure is that it provides no estimate of the error
due to omitting the excited states from the fit.

The constrained curve fitting method of \cite{Lepage:2001ym},
by using Bayesian ideas,
allows one to incorporate the uncertainties due to poorly 
constrained states by relaxing the assumption that there are only a few
states which saturate the correlation function.  
Bayesian fits maximize the probability that the fit function
describes the given data, written as $\prob(f(t;{\bm \lambda})|C(t),I)$;
this probability is related to the likelihood (\ref{eq:likelihood})
by Bayes' theorem
\beq
\prob(f(t;{\bm \lambda})|C(t),I) ~=~ \prob(f(t;{\bm \lambda})|I)  \;
\frac{\prob(C(t)|f(t;{\bm \lambda}),I)}{\prob(C(t)|I)} 
\label{eq:bayestheorem}
\eeq
and is called the posterior probability.
The denominator in (\ref{eq:bayestheorem}) 
is treated as a normalization and plays no
role in finding an optimal set of fit parameters.  On the other hand,
the prefactor, $\prob(f(t;{\bm \lambda})|I)$,
which multiplies the likelihood is the prior probability;
its inclusion is what permits fits to many parameters.

The prior probability contains whatever assumptions
about the values of the fit parameters one can safely make without
looking at the data.
In our case of fitting meson correlators, before any fitting is done
one has an idea of a range of possible values for the
amplitudes $A_k$ and energies $E_k$.  Given such a range,
the least informative prior distribution is a Gaussian
with mean $\mu$ and half-width $\delta$, in which case the prior
probability is given by
\beq
\prob(f(t;\VEC{\lambda})|I) ~=~ \prod_{i=0}^{2K-1} 
\frac{1}{\delta_{\lambda_i}\sqrt{2\pi}} \; \exp \left( 
\frac{(\lambda_i - \mu_{\lambda_i})^2}{2\delta_{\lambda_i}^2} \right)
\label{eq:gaussprior}
\eeq
We sometimes refer to the
set of $\{\mu_i, \delta_i\}$ as the ``priors.''
The quantity which is minimized in the fits is
\bea
\label{eq:chisqaug2}
\chisqaug(C(t),\{\lambda_i\},\{(\mu_i,\delta_i)\}) 
&\equiv& \chi^2(C(t),\{\lambda_i\}) ~+~ \sum_{i=0}^{2K-1} 
\frac{(\lambda_i - \mu_{\lambda_i})^2}{\delta_{\lambda_i}^2} \\
&\propto& -2\ln\prob(f(t;{\bm \lambda})|C(t),I) \nonumber
\; .
\eea

Expression (\ref{eq:chisqaug2}) highlights how the prior distribution
stabilizes the minimization algorithm.
As one increases the number of fit parameters
the terms from the prior in (\ref{eq:chisqaug2}) give curvature to
$\chisqaug$ which prevents the minimization algorithm from
spending much time exploring the flat directions of $\chi^2$ in order to
find a minimum.  The trick now is to distinguish which parameters
are constrained by the data and which ones are fixed by the priors.

A remark on counting the net degrees-of-freedom (DoF) of the
fit is in order.  
As usual each of the data points represents one DoF,
but then each parameter of the fit which is constrained by the
data uses up one of those degrees.  However, in the Bayesian curve fitting
method, there are several fit parameters which are unconstrained
by the data and do not count against the net DoF.  Usually there
are a few parameters obviously constrained by the data and a few
obviously determined solely by the prior, but there may be some
parameters for which such a distinction is not clear.  Therefore,
we simply take the DoF to be the number of data points; instead of
striving for a fit which produces $\chi^2/{\rm DoF} \le 1$, we look
for $\chisqaugdof \approx 1$ together with the 
property that the ratio stays constant as more fit terms are added.

Given a sample of $N$ measurements, one bootstrap
sample is obtained by selecting $N$ measurements, allowing repetitions,
from the original $N$ measurements. 
In principle one would perform a fit on every possible bootstrap sample
generating a Gaussian distribution of bootstrapped fit parameters
$\{\bm{\lambda}^b\}$, the half-width of which gives the bootstrap
uncertainties $\sigma_{\lambda_i}^B$.  However, there are a total
of $(2N-1)!/(N!(N-1)!)\sim N^N$ ways to make a bootstrap 
    sample,\footnote{Counting the
    number of possible bootstraps is equivalent to counting the number of
    ways $n$ indistinguishable balls can be distributed into $k$
    distinguishable buckets: each bucket represents an original
    measurement and the number of balls in a bucket indicates the number
    of times the measurement appears in a given bootstrap sample (and $n=k=N$).
    The answer is called the integer composition of $n$ into $k$ parts and is
    equal to the binomial coefficient 
    $\left(\begin{array}{c}n+k-1\\k-1\end{array}\right)$.}
so it is impossible to generate the entire bootstrap
ensemble -- it also unnecessary.  The bootstrap distribution
can be reliably estimated by randomly generating $N_B$ bootstrap
samples for large enough $N_B$.
We use $N_B\approx N = 200$, and have check that changing $N_B$
by a factor of 2 makes no significant difference in 
$\sigma_{\lambda_i}^B$.

In the unconstrained fitting method $\chi^2$ would be minimized
for each bootstrap sample, resulting in a set of fit parameters
which reproduce the likelihood probability distribution 
(\ref{eq:likelihood}).  For the constrained fits where
we minimize $\chisqaug$, however, it is not enough to resample the 
likelihood -- one must resample the whole posterior distribution, i.e.\
the product of the likelihood and the prior distribution.
Therefore, for each bootstrap sample we randomly choose a new set of
prior means $\{\mu_{\lambda_i}^b\}, \,b\in[1,N_B]$ using the
same distribution used in (\ref{eq:chisqaug2})
\beq
\prob(\mu_{\lambda_i}^b) ~=~ \frac{1}{\delta_{\lambda_i}\sqrt{2\pi}}
\exp\left(-\frac{(\mu_{\lambda_i}^b - \mu_{\lambda_i})^2}
{2\delta_{\lambda_i}^2}\right) \,.
\eeq
The bootstrap fits then
yield an ensemble of $N_B$ results for each fit parameter with
a nearly Gaussian shape
\beq
\prob(\lambda_i^b) ~\sim~ 
\exp\left( -\frac{(\lambda_i^b - \langle\lambda_i\rangle_B)^2}
{2(\sigma_{\lambda_i}^B)^2} \right)
\eeq
where $\lambda_i^b$ is the fit result for $\lambda_i$ on the $b$-th
bootstrap sample, 
$\langle\lambda_i\rangle_B\equiv \sum_{b}^{N_B} \lambda_i^b/N_B$,
and $\sigma_{\lambda_i}^B$ is the bootstrap uncertainty for
$\lambda_i$.  In practice one finds that the distribution of fit
results is Gaussian shaped in the center but has stretched tails
which artificially inflate the quantity $\sqrt{ \langle \lambda_i \rangle_B^2
-  \langle \lambda_i^2 \rangle_B}$  
making it a poor estimate of $\sigma_{\lambda_i}^B$
\cite{Gupta:1987zc,Chu:1991ps}.  Instead
we estimate the width of the bootstrap distribution
by discarding the highest 16\% and lowest 16\% of $\lambda_i^b$ and
quoting the range of values for the remaining 68\% as $2\sigma_{\lambda_i}^B$.
Having obtained bootstrap fits to several correlators, say 
$\{\lambda_i^b\}$ and $\{\nu_j^b\}$, we estimate the uncertainty
in functions of the fit parameters $g(\lambda_i,\nu_j)$, 
for example mass ratios, by computing the function for each 
bootstrap sample and truncating the resulting distribution just as 
discussed for individual fit parameters.

\bibliography{mbw}
\bibliographystyle{apsrev}

\clearpage

\begin{table}
\caption{\label{tab:summary1232} Summary of quenched results from
the isotropic $12^3\times 32$ lattice ($1/a=1.0$ GeV).
Results are checks of our new formulation, not state-of-the-art
computations to be used for phenomenology.}
\begin{ruledtabular}
\begin{tabular}{ll}
\mc{1}{c}{quantity} & \mc{1}{c}{result} \\ \hline
$M_{\rm kin}(B_s)$ & $5.56 \pm 0.33$ GeV \\
$M_{\rm pert}(B_s)$ & $5.51 \pm 0.45$ GeV \\ 
$M_{\rm kin}(B_s^*)$ & $5.68 \pm 0.54$ GeV \\
$M_{\rm pert}(B_s^*)$ & $5.53 \pm 0.45$ GeV \\ 
\\
$B_s^* - B_s$ splitting & $25.0\pm 4.8$ MeV \\
\\
$f_{B_s}$ & $225 \pm 9 {\rm (stat)} \pm 20 {\rm (p.t.)}
\pm 27 {\rm (disc.)}~{\rm MeV}$ 
\end{tabular}
\end{ruledtabular}
\end{table}

\begin{table}
\caption{\label{tab:runlist} Simulation parameters for the quenched
gauge configurations.
There are 200 configurations for each parameter set.}
\begin{center}
\begin{ruledtabular}
\begin{tabular}{rccclllc}
\mc{1}{c}{volume} & \mc{1}{c}{$\beta$} & \mc{1}{c}{$\xi_0$} & $1/a_s$ (GeV) 
& \mc{1}{c}{$a_s/a_t$} & \mc{1}{c}{$u_s$} & \mc{1}{c}{$u_t$} & $a_s M_0$
 \\ \hline
$8^3\times 20$ & $1.719$ & -- & $0.8$ & $1$ & $0.797$ & $0.797$ 
& 6.5  \\
$8^3\times 48$ & $1.8$ & $6.0$ & $0.7$ & $5.3$ & $0.721$ & $0.992$ 
& 7.0  \\
$12^3\times 32$ & $2.131$ & -- & $1.0$ & $1$ & $0.836$ & $0.836$ 
& 5.0  \\
$12^3\times 48$ & $ 2.4$ & $3.0$ & $1.2$ & $2.71$ & $0.7868$ & $0.9771$ & 4.0
\end{tabular}
\end{ruledtabular}
\end{center}
\end{table}

\begin{table}
\caption{\label{tab:priors}
Gaussian prior means $\mu$ and widths $\delta$
for fits to pseudoscalar
heavy-light propagator on the $8^3\times 20$ lattice, $m=0.18$.
We use $e_k$ to denote $E_k$ for the ground states ($k=0,1$) and
$\Delta E_k$ for the excited states ($k\ge 2$).}
\begin{center}
\begin{ruledtabular}
\begin{tabular}{ccc}
$k$ & $\mu_{A_k}\pm \delta_{A_k}$ & $\mu_{e_k}\pm \delta_{e_k}$ \\ \hline
0 & $0.94\pm 0.47$ & $0.900\pm 0.225$ \\
1 & $0.94\pm 0.47$ & $1.40\pm 0.35$\\
2 & $0.60\pm 0.60$ & $0.40\pm 0.30$\\
3 & $0.60\pm 0.60$  & $0.40\pm 0.30$\\
4 & $0.60\pm 0.60$  & $0.40\pm 0.30$\\
5 & $0.60\pm 0.60$  & $0.40\pm 0.30$\\
6 & $0.60\pm 0.60$  & $0.40\pm 0.30$\\
6 & $0.60\pm 0.60$  & $0.40\pm 0.30$
\end{tabular}
\end{ruledtabular}
\end{center}
\end{table}

\begin{table}
\caption{\label{tab:efitvsk}
Dependence of fit results on number of terms ($K$) 
included in fit -- energies of the $8^3\times 20$ heavy-light 
pseudoscalar correlator.  Uncertainties quoted here were estimated
from $\bm{\nabla\nabla}\chisqaug$ as described in the text.}
\begin{center}
\begin{ruledtabular}
\begin{tabular}{ccccc}
\multicolumn{5}{c}{Non-oscillating terms} \\
$K$ & $aE_0$ & $a\Delta E_2$ & $a\Delta E_4$ & $a\Delta E_6$ \\ \hline
   2  &  1.044(0.003) \\
   3  &  0.919(0.013) &  0.492(0.048) \\
   4  &  0.921(0.013) &  0.505(0.061) \\
   5  &  0.915(0.018) &  0.365(0.161) &  0.282(0.250) \\
   6  &  0.917(0.018) &  0.372(0.165) &  0.304(0.259) \\
   7  &  0.914(0.022) &  0.322(0.199) &  0.269(0.257) &  0.348(0.297) \\
   8  &  0.915(0.021) &  0.328(0.203) &  0.288(0.261) &  0.361(0.297) 
\\[.2cm] \hline
\multicolumn{5}{c}{Oscillating terms} \\
$K$ & $aE_1$ & $a\Delta E_3$ & $a\Delta E_5$ & $a\Delta E_7$ \\ \hline
   2 &   1.290(0.015)  \\
   3 &   1.503(0.028)  \\
   4 &   1.470(0.099) &  0.405(0.296) \\
   5 &   1.461(0.103) &  0.422(0.293) \\
   6 &   1.461(0.100) &  0.412(0.299) &  0.412(0.299) \\
   7 &   1.455(0.102) &  0.419(0.299) &  0.419(0.298) \\
   8 &   1.464(0.098) &  0.419(0.300) &  0.418(0.300) &  0.418(0.300) 
\end{tabular}
\end{ruledtabular}
\end{center}
\end{table}

\begin{table}
\caption{\label{tab:afitvsk}
Dependence of fit results on number of terms ($K$) 
included in fit -- amplitudes of the $8^3\times 20$ heavy-light 
pseudoscalar correlator. Uncertainties quoted here were estimated
from $\bm{\nabla\nabla}\chisqaug$ as described in the text. }
\begin{center}
\begin{ruledtabular}
\begin{tabular}{ccccc}
\multicolumn{5}{c}{Non-oscillating terms} \\
$K$ & $A_0$ & $A_2$ & $A_4$ & $A_6$ \\ \hline
   2  & 1.955(0.010) \\
   3  & 1.047(0.104) & 1.088(0.093) \\
   4  & 1.063(0.111) & 1.082(0.091) \\
   5  & 0.999(0.174) & 0.599(0.411) & 0.557(0.411) \\
   6  & 1.014(0.176) & 0.601(0.406) & 0.557(0.409) \\
   7  & 0.980(0.224) & 0.518(0.437) & 0.505(0.456) & 0.185(0.445) \\
   8  & 0.993(0.225) & 0.527(0.432) & 0.488(0.457) & 0.205(0.452)
\\[0.2cm] \hline
\multicolumn{5}{c}{Oscillating terms} \\
$K$ & $A_1$ & $A_3$ & $A_5$ & $A_7$ \\ \hline
   2  & 0.478(0.013) \\
   3  & 0.674(0.024) \\
   4  & 0.580(0.263) & 0.105(0.289) \\
   5  & 0.553(0.265) & 0.140(0.293) \\
   6  & 0.583(0.268) & 0.008(0.448)  & 0.119(0.324) \\
   7  & 0.572(0.267) & -0.005(0.448) & 0.159(0.336) \\
   8  & 0.611(0.275) & -0.043(0.460) & 0.005(0.480) & 0.179(0.412)
\end{tabular}
\end{ruledtabular}
\end{center}
\end{table}

\begin{table}
\caption{\label{tab:chisqaug}
Augmented chi-squared per degree-of-freedom
for the fits in the preceding 2 tables.}
\begin{center}
\begin{ruledtabular}
\begin{tabular}{cccc}
& $K$ & $\chisqaugdof$ & \\ \hline
&   2  &  47.7 & \\
&   3  &  0.60 & \\
&   4  &  0.69 & \\
&   5  &  0.66 & \\
&   6  &  0.74 & \\
&   7  &  0.83 & \\
&   8  &  0.92 &
\end{tabular}
\end{ruledtabular}
\end{center}
\end{table}

\begin{table}
\begin{ruledtabular}
\caption{\label{tab:chisqs}Summary of fits to pseudoscalar
heavy-light correlators. (``AsqTad'' implies $\tnaik=1$ unless
otherwise indicated.)}
\begin{tabular}{dlcccdc}
\mc{1}{c}{$\beta$} & action & $1/a_s$ (GeV) & $1/a_t$ (GeV) 
& $K$ & \mc{1}{c}{$\chisqaugdof$} & $E_{\rm sim}({\bf p}=0)$ (MeV) \\ \hline
1.719 & 1-link & 0.8 & 0.8 & 3 & 0.59 & 735(10) \\ 
1.719 & AsqTad & 0.8 & 0.8 & 3 & 8.93 & -- \\ 
1.719 & Naik & 0.8 & 0.8 & 3 & 17.6 & -- \\ 
1.719 & Fat-link & 0.8 & 0.8 & 3 & 0.51 & 691(20) \\ 
\\
1.8 & AsqTad & 0.7 & 3.7 & 4 & 1.59 & 790(36) \\ 
1.8 & AsqTad ($t_{\rm min}=3$) & 0.7 & 3.7 & 4 & 0.87 & 791(39) \\ 
1.8 & AsqTad-tn & 0.7 & 3.7 & 4 & 1.03 & 901(19) \\ 
\\
2.131 & 1-link & 1.0 & 1.0 & 4 & 0.48 & 873(9) \\ 
2.131 & AsqTad & 1.0 & 1.0 & 4 & 0.96 & 765(9) \\ 
\end{tabular}
\end{ruledtabular}
\end{table}

\begin{table}
\caption{\label{tab:NRKS1link_mesps_820b1719m18}Bootstrap fit results
for the $8^3\times 20$ heavy-light pseudoscalar correlator for
fits to $K$ terms.}
\begin{ruledtabular}
\begin{tabular}{rrrr}
\mc{1}{c}{$\lambda$} & \mc{1}{c}{$K=3$} & \mc{1}{c}{$K=4$} & \mc{1}{c}{$K=5$} \\ \hline 
$A_0$ & 1.043(0.116) & 1.061(0.116) & 1.003(0.183) \\
$aE_0$ & 0.919(0.014) & 0.920(0.014) & 0.918(0.019) \\
$A_1$ & 0.680(0.025) & 0.559(0.270) & 0.538(0.262) \\
$aE_1$ & 1.508(0.030) & 1.457(0.113) & 1.450(0.112) \\
$A_2$ & 1.098(0.102) & 1.098(0.101) & 0.736(0.329) \\
$a\Delta E_2$ & 0.499(0.057) & 0.514(0.070) & 0.405(0.127) \\
$A_3$ &   & 0.141(0.313) & 0.178(0.297) \\
$a\Delta E_3$ &   & 0.442(0.288) & 0.447(0.305) \\
$A_4$ &   &   & 0.496(0.421) \\
$a\Delta E_4$ &   &   & 0.380(0.247) \\
\end{tabular}

\end{ruledtabular}
\end{table}

\clearpage

\begin{table}
\caption{\label{tab:pirho}
Light pseudoscalar and vector meson mass, computed with the same
light quark propagators used for heavy-light mesons.
(``AsqTad'' implies $\tnaik=1$ unless otherwise indicated.)
For comparison, we nominally associate the physical strange
sector with $m_{PS}/m_V = 0.673$.}
\begin{ruledtabular}
\begin{tabular}{rrrrrrr}
\mc{1}{c}{$\beta$} & \mc{1}{c}{$1/a_s$ (GeV)} & \mc{1}{c}{action} & \mc{1}{c}{$a_t m$} & \mc{1}{c}{$m_{PS}$ (MeV)} & \mc{1}{c}{$m_V$ (MeV)} & \mc{1}{c}{$m_{PS}/m_V$} \\ \hline 
1.8 & 0.7 & 1-link & 0.04 & 843(6) & 1251(11) & 0.674(0.005) \\
1.8 & 0.7 & AsqTad & 0.04 & 626(18) & 989(31) & 0.632(0.022) \\
1.8 & 0.7 & AsqTad-tn & 0.04 & 628(19) & 994(37) & 0.630(0.024) \\
1.719 & 0.8 & 1-link & 0.18 & 761(1) & 1171(30) & 0.649(0.017) \\
2.131 & 1.0 & 1-link & 0.12 & 825(2) & 1218(27) & 0.678(0.015) \\
2.131 & 1.0 & AsqTad & 0.10 & 685(3) & 1035(23) & 0.662(0.016) \\
2.4 & 1.2 & 1-link & 0.03 & 808(6) & 1108(21) & 0.728(0.012) \\
2.4 & 1.2 & AsqTad-tn & 0.03 & 619(8) & 913(17) & 0.679(0.013) \\
\end{tabular}

\end{ruledtabular}
\end{table}

\begin{table}
\caption{\label{tab:lspeed_iso}Speed-of-light parameter squared for
several values of $a|\VEC{p}|$ on
the isotropic lattices.  Since no tuning is done, one can estimate
the size of lattice artifacts in finite momentum states from
how different $c^2$ is from 1.}
\begin{ruledtabular}
\begin{tabular}{rrrrrr}
$\beta$ & $L$ & action & \mc{1}{c}{$c^2(2\pi/L)$}
 & \mc{1}{c}{$c^2(2\sqrt{2}\pi/L)$} & \mc{1}{c}{$c^2(2\sqrt{3}\pi/L)$}
 \\ \hline 
1.719 & 8 & 1-link & 0.656(6) & 0.631(8) & -- \\
1.719 & 8 & fat-link & 0.729(17) & 0.712(20) & 0.684(24) \\
1.719 & 8 & Naik & 0.883(12) & 0.916(22) & -- \\
1.719 & 8 & AsqTad & 0.892(14) & 0.922(22) & -- \\
2.131 & 12 & 1-link & 0.794(8) & 0.767(26) & 0.778(16) \\
2.131 & 12 & AsqTad & 0.946(14) & 0.952(26) & 0.836(80) \\
\end{tabular}

\end{ruledtabular}
\end{table}

\begin{table}
\caption{\label{tab:lspeed_aniso}Speed-of-light parameter squared for
several values of $a|\VEC{p}|$
on the anisotropic lattices.  The bare parameter $c_0$ in the
anisotropic action (\ref{eq:anisoasqtad}) is tuned
so that $c^2 \approx 1$.}
\begin{ruledtabular}
\begin{tabular}{rrrrrrr}
$\beta$ & $L$ & action & $c_0$ & \mc{1}{c}{$c^2(2\pi/L)$}
 & \mc{1}{c}{$c^2(2\sqrt{2}\pi/L)$} & \mc{1}{c}{$c^2(2\sqrt{3}\pi/L)$}
 \\ \hline 
1.8 & 8 & 1-link & 1.1 & 1.004(23) & 0.993(31) & 0.991(46) \\
1.8 & 8 & AsqTad & 1.4 & 0.940(46) & 0.952(56) & 0.975(48) \\
1.8 & 8 & AsqTad-tn & 1.4 & 0.948(54) & 0.957(51) & 0.980(45) \\
2.4 & 12 & 1-link & 1.0 & 0.965(35) & 0.957(51) & 0.937(60) \\
2.4 & 12 & AsqTad-tn & 1.0 & 0.931(44) & 0.957(51) & 0.853(81) \\
\end{tabular}

\end{ruledtabular}
\end{table}


\begin{table}
\caption{\label{tab:NRKSasq_mesps_1232b2131m10M50}Bootstrap fit results
for the $12^3\times 32$ heavy-light pseudoscalar ($B_s$) 
correlator for several momenta. (AsqTad light quark action.)  }
\begin{ruledtabular}
\begin{tabular}{rrrrrr}
\mc{1}{c}{$\lambda$} & \mc{1}{c}{$a|\VEC{p}|=0$} & \mc{1}{c}{$a|\VEC{p}|=2\pi/12$} & \mc{1}{c}{$a|\VEC{p}|=2\sqrt{2}\pi/12$} & \mc{1}{c}{$a|\VEC{p}|=2\sqrt{3}\pi/12$} & \mc{1}{c}{$a|\VEC{p}|=4\pi/12$} \\ \hline 
$A_0$ & 0.129(0.009) & 0.130(0.010) & 0.129(0.011) & 0.126(0.012) & 0.139(0.017) \\
$a_t E_0$ & 0.774(0.008) & 0.799(0.009) & 0.822(0.010) & 0.845(0.011) & 0.872(0.014) \\
$A_1$ & 0.071(0.045) & 0.071(0.045) & 0.071(0.046) & 0.070(0.050) & 0.070(0.053) \\
$a_t E_1$ & 1.298(0.090) & 1.312(0.095) & 1.326(0.095) & 1.343(0.100) & 1.370(0.089) \\
$A_2$ & 1.209(0.014) & 1.214(0.014) & 1.222(0.015) & 1.232(0.016) & 1.218(0.019) \\
$a_t\Delta E_2$ & 0.636(0.010) & 0.613(0.010) & 0.589(0.010) & 0.565(0.010) & 0.540(0.010) \\
$A_3$ & 0.738(0.046) & 0.745(0.047) & 0.752(0.048) & 0.760(0.051) & 0.760(0.052) \\
$a_t\Delta E_3$ & 0.149(0.127) & 0.141(0.133) & 0.135(0.120) & 0.124(0.123) & 0.095(0.114) \\
\end{tabular}

\end{ruledtabular}
\end{table}

\begin{table}
\caption{\label{tab:NRKSasq_mesvk_1232b2131m10M50}Bootstrap fit results
for the $12^3\times 32$ heavy-light vector ($B_s^*$) correlator for
several momenta. (AsqTad light quark action.) }
\begin{ruledtabular}
\begin{tabular}{rrrrrr}
\mc{1}{c}{$\lambda$} & \mc{1}{c}{$a|\VEC{p}|=0$} & \mc{1}{c}{$a|\VEC{p}|=2\pi/12$} & \mc{1}{c}{$a|\VEC{p}|=2\sqrt{2}\pi/12$} & \mc{1}{c}{$a|\VEC{p}|=2\sqrt{3}\pi/12$} & \mc{1}{c}{$a|\VEC{p}|=4\pi/12$} \\ \hline 
$A_0$ & 0.112(0.011) & 0.112(0.013) & 0.109(0.016) & 0.104(0.018) & 0.122(0.022) \\
$a_t E_0$ & 0.799(0.010) & 0.823(0.011) & 0.846(0.013) & 0.868(0.016) & 0.898(0.019) \\
$A_1$ & 0.070(0.046) & 0.070(0.048) & 0.069(0.049) & 0.070(0.052) & 0.069(0.052) \\
$a_t E_1$ & 1.339(0.103) & 1.348(0.083) & 1.364(0.084) & 1.379(0.093) & 1.394(0.079) \\
$A_2$ & 1.232(0.014) & 1.238(0.015) & 1.247(0.015) & 1.258(0.018) & 1.240(0.022) \\
$a_t\Delta E_2$ & 0.599(0.009) & 0.575(0.009) & 0.552(0.010) & 0.530(0.011) & 0.503(0.010) \\
$A_3$ & 0.747(0.047) & 0.754(0.049) & 0.761(0.051) & 0.767(0.052) & 0.767(0.052) \\
$a_t\Delta E_3$ & 0.116(0.121) & 0.113(0.110) & 0.105(0.107) & 0.096(0.101) & 0.078(0.092) \\
\end{tabular}

\end{ruledtabular}
\end{table}



\begin{table}
\begin{center}
\caption{\label{tab:Bs_splittings} Mass splittings in the $B_s$ 
spectrum, converted to MeV using $1/a_t$ from Table \ref{tab:runlist}.
The bar over $B_s$ indicates the spin-averaged mass $(M_{B_s}+3M_{B_s^*})/4$
was used.}
\begin{ruledtabular}
\begin{tabular}{rrrrrrrrr}
\mc{1}{c}{$\beta$} & \mc{1}{c}{$1/a_s$ (GeV)} & \mc{1}{c}{action} & $K$ & \mc{1}{c}{$B^*_s-B_s$}
 & \mc{1}{c}{$B^*_{s0}-B_s$}
 & \mc{1}{c}{$B_{s1}-B^*_{s0}$} & \mc{1}{c}{$B^*_{s0}-\overline{B}_s$}
 & \mc{1}{c}{$B_{s1}-\overline{B}_s$} \\ \hline 
1.8 & 0.7 & 1-link & 5 & 34.9(10.2) & 442(56) & 12.6(4.7) & 416(54) & 430(57) \\
1.8 & 0.7 & AsqTad-tn & 4 & 31.9(2.4) & 285(78) & 10.1(3.4) & 261(80) & 272(77) \\
1.719 & 0.8 & 1-link & 3 & 21.1(2.7) & 471(25) & 0.8(2.8) & 456(25) & 456(26) \\
2.131 & 1.0 & 1-link & 4 & 30.7(3.5) & 315(105) & 23.1(24.6) & 292(107) & 321(130) \\
2.131 & 1.0 & AsqTad & 4 & 25.0(4.8) & 523(94) & 35.0(36.1) & 504(92) & 545(101) \\
2.4 & 1.2 & 1-link & 6 & 25.6(12.1) & 425(60) & -9.4(22.2) & 406(57) & 398(63) \\
2.4 & 1.2 & AsqTad-tn & 6 & 32.4(8.0) & 403(56) & 14.2(21.7) & 380(60) & 392(66) \\
\end{tabular}

\end{ruledtabular}
\end{center}
\end{table}

\clearpage

\clearpage

\clearpage

\clearpage

\clearpage

\clearpage

\clearpage

\end{document}